\documentclass[12pt]{article}
\usepackage[T1]{fontenc}

\usepackage{amssymb,amsmath}
\usepackage{graphics}
\usepackage{subfig, graphicx, caption}
\graphicspath{{images/}} 
\usepackage{color}
\usepackage{array,multirow}
\usepackage{epsfig}
\usepackage{mwe}
\usepackage{comment}
\usepackage[a4paper]{geometry}
\usepackage{cite}
\usepackage{placeins}
\usepackage[utf8]{inputenc}
\usepackage{cancel}
\usepackage{arydshln}
\makeatletter \renewcommand{\@dotsep}{10000} \makeatother
\usepackage{indentfirst}
\usepackage[colorlinks=true
,urlcolor=blue
,citecolor=blue
,linkcolor=blue
,pagecolor=blue
,linktocpage=true
,pdfproducer=medialab
]{hyperref}
\def\beq{\begin{equation}}
\def\eeq{\end{equation}}
\usepackage[nodisplayskipstretch]{setspace}

\begin{document}

\begin{titlepage}

\begin{flushright}

\end{flushright}
\pagestyle{empty}

\begin{center}
{\Large \bf $t\bar{t}H$ Interactions and T-odd Correlations at Hadron Colliders}\\
\vspace{8pt}
{\bf  Apurba Tiwari\footnote{\tt E-mail: atiwari@myamu.ac.in} and Sudhir Kumar Gupta\footnote{\tt E-mail:sudhir.ph@amu.ac.in} }
\vspace{2pt}
\begin{flushleft}
{\em Department of Physics, Aligarh Muslim University, Aligarh, UP--202002, INDIA} 
\end{flushleft}

\vspace{10pt}
\begin{abstract}

We explore $\mathcal CP$-violation effects of Higgs-top interactions in the associated production of a Higgs boson with top-pair in 
the dileptonic decay modes of top-quark originating from proton proton collisions $pp \to t\bar{t}H \to (l^+ \nu_l b)(l^{-} 
\bar{\nu_l} \bar{b})H$ at NLO in QCD matched to parton shower via T-odd observables using momenta of various particles involved in the 
process. In particular, we predict the constraints on the $\mathcal CP$-violating $t\bar{t}H$ coupling obtained through the production 
asymmetries associated with the T-odd observables in the dileptonic decay channel of the $t\bar{t}$ pair for the LHC with 
centre-of-mass energy of 13 TeV and an integrated luminosity of 139 fb$^{-1}$. We also present the corresponding limits for future 
hadron colliders, namely the High Luminosity LHC (HL-LHC) and the Future Circular Collider (FCC-HH). Our estimates of the $t\bar{t}H$ 
interaction strength reveal that the upper bound on pseudoscalar coupling $c_p$ corresponding to the largest asymmetry would be of 
about $1.96 \times 10^{-2}$ at 2.5$\sigma$ C.L. for $c_s$ = 1 at the LHC with $\sqrt{S}$ = 13 TeV for an integrated luminosity of 139 
fb$^{-1}$. The respective limits for HL-LHC and FCC-hh with the projected Luminosities of 3 ab$^{-1}$ and 30 ab$^{-1}$ are found to be
to $3.4\times 10^{-3}$ and $1.6\times 10^{-4}$ respectively at 2.5$\sigma$ C.L.

\end{abstract} 
\end{center}

\end{titlepage}

\section{Introduction} 
\label{intro} 

The Standard Model (SM) \cite{Novaes:1999yn,Herrero:1998eq,Langacker:2009my,Kibble:2014spa} was considered an enormously successful 
and affluent model prior to experimental evidence in the discovery of the Higgs boson \cite 
{Dawson:1994ri,Organtini:2012ut,Peskin:2015kka,Bass:2021acr} which is an essential component of SM. The discovery of the Higgs boson 
marked the beginning of a new era in particle physics when the ATLAS \cite{ATLAS:2012yve} and CMS \cite{CMS:2012qbp} in Run-1 of the 
Large Hadron Collider (LHC) strongly confirmed the existence of the Higgs boson with a mass of about 125 GeV, which is analogous to 
the SM Higgs boson with spin-zero and parity even. After the discovery of the Higgs boson, it has become paramount to determine its 
physical properties. Some of its properties like spin and mass have been measured with data sets collected during Run-1 and Run-2 of 
the LHC that are identical to those predicted by SM within the limits of theoretical and experimental uncertainties. Deviations are 
still allowed since the Higgs boson in the SM has a very unnatural mass and the current data show large uncertainties in the 
measurement of many Higgs couplings, such as Higgs-top couplings, indicating that these need to be explored precisely. The fact that 
there are large uncertainties in the measurement of the Higgs boson coupling with the fermion and vector bosons indicates that there 
is ample room for the existence of new physics, therefore the Higgs boson sector is of high relevance for studying new physics effects 
in BSM investigations. Consequently, one of the main objectives of the future LHC run is to investigate the true nature of the Higgs 
boson and precisely measure the Higgs boson properties such as its coupling with SM particles, its $\mathcal CP$ nature, etc. To 
absolutely establish the true nature of the Higgs boson requires precisely measuring the Higgs coupling to fermion and gauge bosons 
and the Higgs self-coupling, so the important task of upcoming experiments at the LHC is to measure these couplings to the greatest 
possible accuracy. The LHC is currently undergoing significant upgrades for its upcoming Run which is Run-3 which will be a High 
Luminosity Phase \cite{Apollinari:2015wtw} where the Luminosity will be significantly enhanced. It is anticipated that much larger 
data sets will be collected during Run-3 which will lead to better understanding of systematic uncertainties and increase experimental 
accuracy by substantially reducing experimental errors. Considering such promising experimental possibilities in future High 
Luminosity and high energy experiments, it is beneficial to conduct a comprehensive study to explore the properties of the Higgs boson 
in different BSM scenarios.

Exploring the $\mathcal CP$ nature of the Higgs boson interactions is crucial in order to find an explanation for the existing 
imbalance between matter and antimatter. Since the SM does not provide a sufficient amount of $\mathcal CP$-violation to explain the 
current matter-antimatter asymmetry of the Universe \cite{Manousakis:2022ecy,Dine:2003ax,Morrissey:2012db,Buchmuller:2005eh}, 
exploring Higgs boson interactions may provide new sources for the investigation of such phenomena in many beyond the Standard Model 
(BSM) theories. Interestingly, new sources of $\mathcal CP$-violation may play a major role in understanding the present asymmetry, 
hence exploring new $\mathcal CP$-violation sources beyond SM theories is of utmost importance for the future Hadron Colliders. In 
addition to unravelling the matter-antimatter asymmetry, studies concerning the hierarchy problem or the naturalness problem 
\cite{Gildener:1976ih,Weinberg:1975gm,Susskind:1978ms}, dark matter (DM) abundance \cite{Jungman:1995df,Sadeghian:2013bga}, 
non-vanishing neutrino masses \cite{Gil-Botella:2013bnb}, inflation \cite{Lyth:1998xn}, etc. specifically demand exploring the Higgs 
boson extensions.

The interaction of the Higgs boson with the heaviest fermion, that is, the top-quark, is phenomenologically and theoretically 
important as it has the strongest coupling with the Higgs boson since the associated Yukawa coupling is the largest. Thus, the 
accurate measurement of the Higgs-top interaction plays a crucial role in establishing the true nature of the Higgs boson and also 
contributes to understanding the vital problem of vacuum stability \cite{Sher:1988mj,Degrassi:2012ry} and several cosmological 
phenomena, such as baryogenesis \cite{Zhang:1994fb,Kobakhidze:2015xlz} and electroweak phase transition \cite{Huang:2015izx} etc. The 
possible production modes of the Higgs boson at the LHC are: gluon-gluon fusion (ggf), vector-boson fusion (VBF), associated 
production with a $W^{\pm}$ or Z boson (VH, V=W or Z) and production in association with a single top-quark ($tH$) or with a top-pair 
($t\bar{t}H$). Although Higgs-top coupling can be accessed through loop-induced processes 
\cite{Ellis:2013lra,Brod:2013cka,ACME:2013pal} however the leading contribution comes primarily through two possible processes: a) the 
production of the Higgs Boson in association with top pair 
\cite{Goldstein:2000bp,Belyaev:2002ua,Drollinger:2001ym,Beenakker:2002nc,Agrawal:2013owa,Biswas:2014hwa,Garzelli:2011vp,Frederix:2011zi,Degrande:2012gr,Curtin:2013zua,Adelman:2013vro, 
Marciano:1991qq,Buckley:2015vsa,Cao:2016wib,Maltoni:2016yxb,Gritsan:2016hjl,Chang:2016mso} and b) associated production of the Higgs 
Boson with single top 
\cite{Bordes:1992jy,Ballestrero:1992bk,Stirling:1992fx,Diaz-Cruz:1991bks,Maltoni:2001hu,Lu:2010zzb,Farina:2012xp,Biswas:2012bd,Ellis:2013yxa,Englert:2014pja,Chang:2014rfa,Wu:2014dba, 
Yang:2014xma,Yue:2014tya,Rindani:2016scj,Liu:2016gsi,Kobakhidze:2014gqa} at the LHC. The production rate of the Higgs boson in 
association with the top pair is relatively high and is dominant, therefore, crucial to potentially disentangle the new physics 
effects \cite{Buckley:2015vsa,Bortolato:2020zcg,Cao:2020hhb,Goncalves:2021dcu,Mileo:2016mxg}. Furthermore, the importance of the 
process $pp \to t\bar{t}H$ \cite{Dawson:2003zu,Frederix:2011zi} lies in the fact that the actual presence of the top-quark can be 
observed in the final state particles \cite{Maltoni:2002jr,Belyaev:2002ua}. In the case of Yukawa coupling, the $\mathcal CP$ -odd 
contribution in the interaction of Higgs to fermion is unsuppressed and therefore the study of the Higgs-fermion-fermion interaction 
is much more useful in this respect and will allow us to have a clear understanding of the $\mathcal CP$ structure of the Higgs Boson. 
The $t\bar{t}H$ channel was detected by ATLAS with a significance of about 6.3 and by CMS with a significance of approximately 5.2 
\cite{CMS:2018uxb,ATLAS:2018mme}.

In the present manuscript, we perform a systematic and detailed investigation of the $\mathcal CP$-violating effects of the Higgs-top 
coupling using T-odd observables considering the dominating Higgs production process, $pp \to t\bar{t}H$ in the dileptonic decay mode 
of the top-quark. We conduct the analysis at Next-to-leading order (NLO) accuracy in QCD matched to parton-showers (NLO+PS). The 
present study aims to explore the potential of the T-odd triple product correlations constructed via the momenta of the final decay 
products in the corresponding Higgs top-pair production at the LHC and Future Hadron Colliders for improving $\mathcal CP$-violation 
sensitivity to anomalous Higgs-top interactions. We work in an effective field theory framework to make our study model-independent 
and analytical enough for studying CP-violating Higgs-top interactions in the BSM models. In this approach, an effective Lagrangian is 
constructed by introducing higher dimensional operators to incorporate new physics contributions to the Standard Model and where the 
$t\bar{t}H$ vertex is parameterized in terms of two unknown factors, a $\mathcal CP$-even component $c_s$ and a $\mathcal CP$-odd 
component $c_p$. Particularly in this work, the main idea is to find the constraints on the $\mathcal CP$-violating anomalous 
Higgs-top coupling at NLO accuracy including parton-shower effects (NLO+PS) from cross-section measurements as well as from production 
asymmetries for the LHC with $\sqrt{S}$ = 13 TeV and an integrated luminosity of $\int L dt$ = 139 fb$^{-1}$. In addition, we derive 
projections for Future Hadron Colliders, namely, HL-LHC and FCC-hh for $\sqrt{S}$ = 14 TeV and 100 TeV with luminosities of 3 
ab$^{-1}$ and 30 ab$^{-1}$, respectively.

So far $\mathcal CP$-violation in Higgs-top interactions via Higgs production and decay has received considerable attention and has 
been extensively probed in the literature 
\cite{Khatibi:2014bsa,Bahl:2021dnc,Ellis:2013yxa,Schlaffer:2014osa,Englert:2014pja,Kobakhidze:2014gqa,Chang:2014rfa,Biswas:2014hwa, 
Goldstein:2000bp,Belyaev:2002ua,Maltoni:2002jr,Drollinger:2001ym,Beenakker:2002nc,Agrawal:2013owa,Garzelli:2011vp, 
Frederix:2011zi,Degrande:2012gr,Curtin:2013zua,Adelman:2013vro,Bordes:1992jy,Ballestrero:1992bk,Stirling:1992fx,Diaz-Cruz:1991bks, 
CMS:2020dkv,CMS:2020cga,ATLAS:2020ior,CMS:2021nnc,ATLAS:2018mme,Nishiwaki:2013cma,Artoisenet:2013vfa,Dai:1993gm,Schmidt:1992et, 
Barman:2021yfh,Bortolato:2020zcg,Bahl:2020wee,Bahl:2022yrs,ATLAS:2023cbt}. However Refs. 
\cite{Hermann:2022vit,Maltoni:2016yxb,Maltoni:2013sma,Demartin:2014fia} discuss the Higgs interactions at NLO accuracy in QCD 
including the parton shower effects. The first phenomenological study of the Higgs production in association with top-antitop pair to 
NLO accuracy in QCD matched to parton showers for both $\mathcal CP$-even and -odd cases was performed in Ref. \cite{Frederix:2011zi}. 
Ref. \cite{Demartin:2014fia} first provided a complete analysis of the Higgs-top interaction in the $t\bar{t}H$ channel at NLO in QCD, 
interfaced with parton showers for BSM studies by including the necessary Ultraviolet (UV) and $R_2$ terms in the UFO model. The model 
is also publicly available. Also, they showed that angular pseudorapidity separation between the leptons or between b-jets in the 
context of the Higgs production with top pair is a promising observable to probe $\mathcal CP$ nature of Higgs interactions at high 
$P_T$ region. In addition, studies related to Higgs coupling to gluons \cite{ATLAS:2021pkb}, tau leptons 
\cite{ATLAS:2015xst,CMS:2017zyp,CMS:2020rpr,CMS:2021sdq}, and muons \cite{ATLAS:2020fzp, CMS:2020xwi} have also been explored 
independently. Besides, $\mathcal CP$-violation in the processes of top-production and decay has also been searched immensely in the 
existing literature 
\cite{Cirigliano:2016nyn,Gupta:2009wu,Tiwari:2019kly,Alioli:2017ces,Aguilar-Saavedra:2008quj,Tiwari:2022nli,CMS:2014uod,Gupta:2009eq}. 
Upper limits on neutron, mercury and electric dipole moments (EDMs) have placed indirect constraints on the anomalous Higgs-top 
interactions. The contribution of the $\mathcal CP$-violation component of the anomalous Higgs-top interactions determines the 
strongest constraint for electron EDM \cite{Ellis:2013lra}. Furthermore, other studies such as low-energy physics probes set 
relatively lower bounds 
\cite{Brod:2013cka,Dolan:2014upa,Englert:2012xt,Kobakhidze:2016mfx,Bernlochner:2018opw,Englert:2019xhk,Gritsan:2020pib,Bahl:2020wee}.

The structure of the paper is as follows: In Section \ref{process&observables}, we discuss the parameterization for the Higgs-top 
Yukawa coupling and present the effective Lagrangian that we have considered as a benchmark model for our study. Furthermore, we 
discuss the key observables relevant for our study to probe the $\mathcal CP$-violation sensitivity of the Higgs top interaction. In 
Section \ref{num_analy}, we present a detailed analysis and investigate the sensitivity of the anomalous Higgs-top coupling through 
cross-section measurements at NLO+PS. Also in the same section, we construct the production asymmetries corresponding to the various 
$\mathcal CP$-violating observables, defined in section \ref{process&observables} in the context of the di-leptonic decay modes of 
both top-quarks and derive constraints on the anomalous $t\bar{t}H$ coupling corresponding to the most promising asymmetries 
encountered here at the LHC and the Future Hadron Colliders such as HL-LHC and FCC-hh. Finally, Section \ref{R&D} concludes the paper.

\section{$pp \to t\bar{t}H$ Process and T-odd Observables}
\label{process&observables}

\begin{figure}[h!]
\centering
\includegraphics[width=0.85\linewidth]{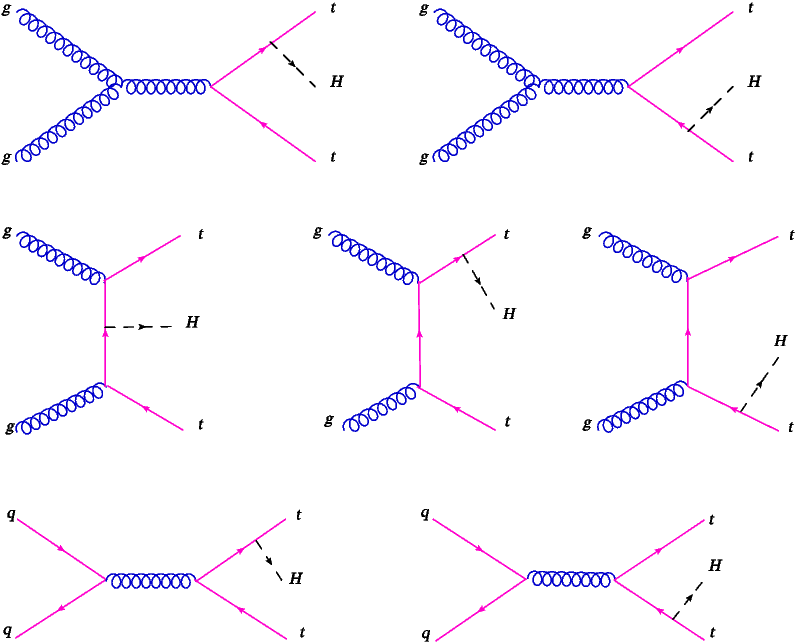}
\caption{\small Representative parton-level Feynman diagrams of the process $pp \to t\bar{t}H$ in leading order at the LHC. The diagrams 
were obtained through $\tt JaxoDraw$ \cite{Binosi:2003yf,Binosi:2008ig}.}
\label{ppttH_process}
\end{figure}

The present study includes the Higgs production in association with a top and anti-top-quark where the top (anti-top)-quark further 
decays leptonically into $bl^{+}\nu_l$ ($\bar{b}l^{-}\bar{\nu_l}$). However, there can be possibilities for many other potential 
signatures, although we are mainly interested only in those signatures in which most of the leptons are in the final state as this 
would lead to a smaller background. Representative parton level diagrams showing the production mechanism of the Higgs boson with the 
top-pair are displayed in Fig. \ref{ppttH_process}. It includes the production of Higgs with top-pair via gluon-gluon annihilation and 
$q\bar{q}$ annihilation, where the first five diagrams in the two rows represent production via gluon-gluon annihilation and the last 
row shows the $q\bar{q}$ annihilation process. The process $pp \to t\bar{t}H$ occurs primarily through the gluon-gluon annihilation 
process $gg \to t\bar{t}H$ while the leading contribution comes from one-loop diagrams. Gluon-initiated processes are anticipated to 
play a vital role in the search for new physics at the Hadron Colliders as gluon luminosity increases with an increase in 
center-of-mass energy. On the other hand, the $q\bar{q}$ annihilation processes, due to Higgs boson coupling and very light quarks, 
make a negligible contribution. The current study aims to investigate the $\mathcal CP$ -violating effects of Higgs-top coupling 
arising due to the presence of the $t\bar{t}H$ vertex in the $pp \to t\bar{t}H$ process. In SM, Higgs-top coupling is a purely scalar 
interaction and consists only scalar-type of components, whereas models beyond SM include both scalar and pseudoscalar couplings as 
the non-linear perception of electroweak gauge symmetry comes into the scenario. The Higgs boson can also be a $\mathcal CP$-mixed 
state in these models \cite{Accomando:2006ga,Koulovassilopoulos:1993pw} and the possibility of a mixture of scalar and pseudoscalar 
couplings has also been confirmed by the present data \cite{Ellis:2013yxa,Nishiwaki:2013cma}. Typically, constructing a model that can 
specifically induce $\mathcal CP$ -violating operators is difficult. Furthermore, the idea of a completely $\mathcal CP$-odd Higgs 
boson has not been favored experimentally and the LHC experiment has already rejected speculation of a pure $\mathcal CP$-odd Higgs. 
However, the constraint imposed by the experiments on the $\mathcal CP$-mix state, that is, mixing of $\mathcal CP$-odd and 
$\mathcal CP$-even states of the Higgs boson, is very low \cite{CMS:2014nkk,CMS:2016tad}. Consequently, a model that contains a 
Lagrangian with both $\mathcal CP$-even and $\mathcal CP$-odd components, will be more realistic.

We consider the following most general parameterization of the Higgs-top Yukawa coupling by employing an effective Lagrangian
\cite{Bahl:2021dnc},
\begin{eqnarray}
\label{Lag_eqn}
L_{t\bar{t}H} &=& - \frac{y_t}{\sqrt{2}} \bar{t}(c_s + i c_p \gamma_5)tH,
\end{eqnarray}

where $H$ is the Higgs field, $y_t$ = $\sqrt{2} \frac{m_t}{v}$ in which $m_t$ is the mass of the top-quark and $v$ is the electroweak 
symmetry breaking scale known as the Higgs vacuum expectation value ($v$ = 246 GeV). $c_s$ and $c_p$ are dimensionless parameters 
known as scalar and pseudo-scalar component that represent $\mathcal CP$-even and $\mathcal CP$-odd anomalous Higgs-top interaction, 
respectively. In SM at the tree level, $c_s$ and $c_p$ take the values $c_s$ = 1 and $c_p$ = 0, while a non zero value of the coupling 
$c_p$ would indicate a deviation towards the non SM and show beyond the SM contribution. It is to be noted here that the $\mathcal 
CP$-violating component $c_p$ receives a contribution from the loop at higher order in the SM but is very small. However, in models 
beyond SM, it is anticipated that $\mathcal CP$-violating component may receive a relatively large contribution from higher order 
effects such as higher dimensional operators in an effective field theory scenario. It is worth mentioning here the importance of 
considering the production of Higgs boson with top-pair which lies in the fact that the top-quark is unique among all quarks owing the 
largest mass $m_t$ = 172.5 $\pm 0.7$ GeV \cite{ParticleDataGroup:2022pth}. It is because of its large mass, the top-quark has a very 
short life-time ($\sim 10^{-25}$ sec), which is shorter than the characteristic hadronization time scale. Therefore it decays rapidly
before it can form any bound state by getting affected with the non-perturbative QCD effects and hence the properties of the top-quark 
are passed on entirely to its decay products. Consequently, the top-quark properties can be traced back from its decay products as 
decay products preserve information of its properties and are therefore useful for investigating direct $\mathcal CP$-violation 
effects in these events.

Several kinematic observables sensitive to the Higgs-top $\mathcal CP$ structure have been discussed in the literature so far, e.g., 
cross-section, invariant mass distribution, transverse Higgs momentum distribution and azimuthal angular separation between $t\bar{t}$
and $tH$ \cite{Demartin:2015uha,Demartin:2014fia,Goncalves:2018agy,Gunion:1996xu}. But these are $\mathcal CP$-even observables and 
require the full reconstruction of the top and the anti-top momenta. Such observables only give an indirect measure of $\mathcal 
CP$-violation and are therefore not helpful in investigating direct $\mathcal CP$-violation. To probe direct $\mathcal CP$-violation, 
$\mathcal CP$-odd observables must be considered. Typical $\mathcal CP$-odd observables can be constructed using the angular 
distribution but these are often experimentally challenging. Alternatively, an indirect approach is considered which can provide 
important complementary information. Although an indirect approach can impose stringent constraints on the $\mathcal CP$-violating 
coupling, it is not necessary that the contribution is solely from $\mathcal CP$-violating interactions and may also be due to 
$\mathcal CP$-even interactions. The genuine $\mathcal CP$-odd observable, as we considered in our study can be constructed from the 
anti-symmetric tensor product defined as $\epsilon (a, b, c, d)$ = $\epsilon_{\mu\nu\alpha\beta}a^{\mu}b^{\nu}c^{\alpha}d^{\beta}$ 
\cite{Hayreter:2015ryk,Gupta:2009wu}. Another prominent example of $\mathcal CP$-odd observables are EDMs that place strong 
constraints on $\mathcal CP$-violating Higgs-top anomalous coupling as we discussed before. As we mentioned earlier, the goal of the 
present study is to investigate the $\mathcal CP$-violation effects of the anomalous Higgs-top interactions in the target process $pp 
\to t\bar{t}H$. For this, several observables can be proposed to investigate $\mathcal CP$-violation effects, particularly in this 
study we consider the T-odd triple product correlations \cite{Tiwari:2019kly,Gupta:2009eq,Gupta:2009pv,Gupta:2009wu} constructed 
through the momenta of the end state products. The T-odd correlations represent the naive T-odd \cite{Han:2009ra} which can also be 
$\mathcal CP$-even and not necessarily $\mathcal CP$-odd. In Ref. \cite{Tiwari:2019kly} many T-odd correlations were identified in 
view of the anomalous top-quark interactions in the process $pp \to t\bar{t} \to (bl^{+}\nu_l) (\bar{b}l^{-}\bar{\nu_l})$. Of these, 
we have projected out some observables that are best suited for our study and can exploit the final decay channel of the target 
process i.e. $pp \to t\bar{t}H \to (bl^{+}\nu_l)(\bar{b}l^{-}\bar{\nu_l})H$. Along with them, we propose other 
possibilities that can explore the final decay state in greater detail and may increase the sensitivity to the anomalous Higgs-top 
coupling. We devote special attention to the inclusion of the observables that contain momenta of Higgs boson to investigate 
the $\mathcal CP$-violation effects that arise due to the presence of the Higgs boson. The analysis considers the following 
observables:
\begin{eqnarray}
\label{observables}
\nonumber
\mathcal O_1  &\equiv&  \epsilon(P,p_b-p_{\bar{b}},p_{l^+},p_{l^-}),\\
\nonumber
\mathcal O_2  &\equiv&  \epsilon(p_h, p_b - p_{\bar{b}}, p_{l^+}, p_{l^-}),\\
\nonumber
\mathcal O_3  &\equiv&  \epsilon(p_b,p_{\bar{b}},p_{l^+},p_{l^-}),\\
\nonumber
\mathcal O_4  &\equiv&  q \cdot (p_{l^+}-p_{l^-})~\epsilon(p_b,p_{\bar{b}},p_{l^+}+p_{l^-}, q),\\
\nonumber
\mathcal O_5  &\equiv&  \epsilon(p_b + p_{l^+},p_{\bar{b}} + p_{l^{-}},p_b+p_{\bar{b}},p_{l^+}-p_{l^-}),\\
\nonumber
\mathcal O_6  &\equiv&  \epsilon(P, p_h, p_b - p_{\bar{b}}, p_{l^+} - p_{l^-}),\\
\mathcal O_7  &\equiv&  \epsilon(q, p_h, p_{b} - p_{\bar{b}}, p_{l^+} - p_{l^-}),
\end{eqnarray}

where $\epsilon$ represents the Levi Civita symbol of rank 4 which is completely anti-symmetric with $\epsilon_{0123} = 1$, which is 
contracted with the four vectors a, b, c, and d as $\epsilon (a, b, c, d)$ = 
$\epsilon_{\mu\nu\alpha\beta}a^{\mu}b^{\nu}c^{\alpha}d^{\beta}$; $p_{l^+}~(p_{l^-})$ denotes the momenta of the 
lepton (anti-lepton) which is identified as arising from the $W^{+}~(W^{-})$ boson and $p_b~(p_{\bar{b}})$, $p_h$ refer the momenta of 
the $b~(\bar{b}$)-quark, Higgs boson respectively. P is defined as the sum of four momenta of b-quark, 
$\bar{b}$-quark, lepton, anti-lepton and Higgs and q is the difference of two beam four momenta, i.e.,
\begin{eqnarray}
\label{P&q}
\nonumber
P &\equiv& p_b + p_{\bar{b}} + p_{l^+} + p_{l^-} + p_h,\\
q &\equiv& P_1 - P_2.
\end{eqnarray}

The observables defined in Eq. \ref{observables} are proportional to the triple product and take the form $\vec p_1.(\vec p_2 \times 
\vec p_3)$, where $\vec p_i~(i =1, 2, 3)$ are momentum vectors. All these observables listed above are odd under $\mathcal 
CP$-transformation and hence these constructions are $\mathcal CP$-odd. Furthermore, there are many other possibilities that we have 
not listed above as the sensitivity corresponding to them is not significant to account for. Before moving on to further analysis, let 
us first discuss some important points related to the observables listed above and discuss how much information is required to conduct 
an efficient analysis. The importance of the above observables lies in the fact that they do not demand spin-related information of 
the produced particles nor do they require the reconstruction of the top-quark, rather they are constructed with momentum of the 
reconstructable final state particles which can be well measured in the LHC experiment. Some of these observables do not need to 
distinguish between $b$ and $\bar{b}$-quark while others require it. However, for the cases where $b$ and $\bar{b}$ 
distinction is required, direction of leptons for the corresponding b-quarks could be used i.e. $b$-jet that will be near $l^{+}$ will 
be identified as originated from $b$-quarks and another $b$-jet closer to $l^{-}$ would have originated from $\bar{b}$-quarks.

Obtaining non-zero value for a $\mathcal CP$-odd observable would be a clear sign of $\mathcal CP$-violation and therefore 
ensures the existence of new physics. Our search strategy relies on the measurement of asymmetry. Therefore, we conduct the analysis
by constructing asymmetries corresponding to each observable given in Eq. \ref{observables} through the following expression:
\begin{eqnarray}
\label{Asymm_form}
\mathcal A_{\mathcal CP}  &=&  {\frac{N(\mathcal O_i>0) - N(\mathcal O_i<0)}{N(\mathcal O_i>0) + N(\mathcal O_i<0)}},
\end{eqnarray}

The above expression for the calculation of $\mathcal CP$-violating asymmetry gives the difference between the number of events for 
which an observable is positive to the number of events for which it is negative, normalised to the total number of events. The 
presence of $\mathcal CP$-violation in the Higgs–top interactions would be manifested by a non-zero value of the asymmetry $\mathcal 
A_{\mathcal CP}$. In SM, the asymmetry $\mathcal A_{\mathcal CP}$ will be negligible for all the observables listed in Eq. 
\ref{observables}. However, in case of beyond SM where anomalous $t\bar{t}H$ coupling comes into the scenario, sizable asymmetries can 
be produced. Since The prime focus of our study is to search for $\mathcal CP$-violating effects of anomalous Higgs-top interactions 
and since a non-zero value of asymmetry $\mathcal A_{CP}$ would indicate the presence of $\mathcal CP$-violation, we will present the 
values of the asymmetries measured as one of the primary results of our analysis that would indeed be the explicit verification of the 
statement given earlier in our analysis that all the observables are $\mathcal CP$-odd.

\section{Numerical Analysis} 
\label{num_analy}

We begin the analysis considering the Higgs characterisation model proposed in Ref. 
\cite{Artoisenet:2013puc,Maltoni:2013sma,Demartin:2014fia} to study the $\mathcal CP$ properties of the Higgs-top interactions in the 
associated Higgs production with top pair at NLO accuracy. We perform the analysis in a fully automatic manner by incorporating the 
relevant interaction Lagrangian given in Eq. \ref{Lag_eqn} in $\tt FeynRules$ \cite{Alloul:2013bka,Christensen:2008py}. The resulting 
UFO model was used to generate $t\bar{t}H$ events at NLO accuracy with the aid of the $\tt MadGraph5\_aMC@NLO$ 
\cite{Alwall:2011uj,Frederix:2009yq,Alwall:2014hca,Hirschi:2011pa} framework and then the produced events were passed to $\tt Madspin$ 
\cite{Artoisenet:2012st} for the top (anti-top)-quark to decay into $bl^{+} \nu_l~(\bar{b}l^{-} \bar{\nu_l})$. The decayed events were 
then interfaced to $\tt Pyhtia8$ \cite{Sjostrand:2014zea,Bierlich:2022pfr} for parton showering and Hadronization. Experimental values 
of the SM input parameters considered in our study are presented in Table \ref{SMinputs}, the central values for renormalization and 
factorization scale has been set to $M_Z$, the parton density function (PDF) has been considered to be NN23NLO 
\cite{Ball:2013hta,NNPDF:2014otw} and the strong coupling constant has been set to a value of $\alpha_s = 0.118$. We conduct the 
analysis at LHC with $\sqrt{S}$ = 13 TeV and an integrated luminosity of $\int L dt$ = 139 fb$^{-1}$. Moreover, we explore the Future 
Hadron Colliders, viz, HL-LHC and FCC-hh with $\sqrt{S}$ = 14 TeV and 100 TeV, respectively.

The effects of the new physics contribution in Eq. \ref{Lag_eqn} will be reflected in Higgs production as well as Higgs decay. 
Although our study aims to investigate new physics effects in Higgs production, exploring final state objects, as the total signal 
rate changes more relevantly with Higgs decays. This would uplift new physics sensitivity from the $\mathcal CP$ violating effects 
of the Higgs-top Yukawa interactions.
\begin{table}[!ht]
\centering
\scalebox{1.0}{ 
\renewcommand{\arraystretch}{1.4}
\begin{tabular} {l|l}
\hline
SM parameter & Experimental value\\
\hline $m_b (m_b)$ & $4.18^{+0.03}_{-0.02}$ GeV \\
$m_t(m_t)$ & 172.5 $\pm$ 0.7 GeV \\
$M_W$ & 80.377 $\pm$ 0.012 GeV \\
$M_Z$ & 91.188 $\pm$ 0.0021 GeV \\
$M_H$ & 125.25 $\pm$ 0.17 GeV \\
\hline
\end{tabular}}
\caption{Experimental values of Standard Model input parameters \cite{ParticleDataGroup:2022pth}.}
\label{SMinputs}
\end{table}

We impose the kinematic cuts on the final state objects by closely following the experimental analysis, which corresponds to a 
simplified version of the standard pre-selection cuts used by the CMS experiment for the measurement of the $t\bar{t}H$ channel 
\cite{CMS:2013szn}. Our choice of the event selection criteria resembles the basic cuts used in Ref. \cite{Khatibi:2014bsa}. All 
events require objects with pseudorapidity $\left|\eta\right|$ < 2.5 and the transverse momentum $P_T$ > 25 GeV. Besides the angular 
separation between the objects is assumed to be $R$ > 0.4.

Using the setup defined above, we generate 10 million events for 10 different benchmark points. We allow to vary both $c_s$ and $c_p$ 
to different values. Specifically, we examine the cases $c_p$ = 0.0, 0.2, 0.4, 0.6, 0.8 and 1.0 and $c_s$ = 0.8, 0.9 and 1.0. Among 
these benchmark points, the point corresponding to $c_s$ = 1 and $c_p$ = 0 represents the SM. After event generation and imposing the 
cuts, we calculate the asymmetry for the set of observables listed in Eq. \ref{observables} for LHC with $\sqrt{S}$ = 13 TeV, HL-LHC 
with $\sqrt{S}$ = 14 TeV and FCC-hh with $\sqrt{S}$ = 100 TeV. The results obtained are then used for further simulation.

The results obtained for the asymmetries at Next-to-leading order interfaced to parton showers (NLO+PS) for the set of observables 
defined in Eq. \ref{observables} for LHC with $\sqrt S$= 13 TeV, HL-LHC with $\sqrt{S}$ = 14 TeV and FCC-hh with $\sqrt{S}$ = 100 TeV 
are shown in Tables \ref{Asym_13TeV_NLO_PS}, \ref{Asym_14TeV_NLO_PS} and \ref{Asym_100TeV_NLO_PS}, respectively. It is to be noted 
that the signal should be larger than 3$\sigma$ C.L. i.e. $\mathcal A_i \sim 1 \times 10^{-3}~(0.1 \%)$. According to Tables 
\ref{Asym_13TeV_NLO_PS}, \ref{Asym_14TeV_NLO_PS} and \ref{Asym_100TeV_NLO_PS}, almost all observables except $\mathcal O_7$ are found 
to be non-zero at 3$\sigma$ C.L., although some observables are only weakly sensitive to the $\mathcal {CP}$-violating coupling $c_p$ 
of the Higgs-top interaction. The observables $\mathcal O_1$, $\mathcal O_2$, $\mathcal O_3$, $\mathcal O_4$, $\mathcal O_5$ and 
$\mathcal O_6$ give promising results as the asymmetries corresponding to them are non zero within the statistical uncertainty at 
3$\sigma$ C.L. and significantly large. Additionally, the observable $\mathcal O_2$ is quite interesting to explore the $\mathcal CP$ 
effects as it involves Higgs momentum as well. We will therefore find constraints on the pseudo-scalar coupling $c_p$ corresponding to 
the observables $\mathcal O_1$, $\mathcal O_2$, $\mathcal O_3$, $\mathcal O_4$, $\mathcal O_5$ and $\mathcal O_6$ that are strongly 
sensitive to the $\mathcal CP$-odd nature of the Higgs-top interaction and will not pursue for the observable $\mathcal O_7$. 
Consequently, we expect promising results as the asymmetries corresponding to these observables are large. From Tables 
\ref{Asym_13TeV_NLO_PS}, \ref{Asym_14TeV_NLO_PS} and \ref{Asym_100TeV_NLO_PS} two comments are in order. First, we see that 
asymmetries become large as we increase the value of pseudo-scalar coupling $c_p$. Second, all the observables $\mathcal O_{1-7}$ do 
not get contributions from the SM ($c_s$ = 1, $c_p$ = 0), confirming our a-priori statement that the asymmetries in the SM are zero. 
Our results are based only on phenomenological simulations and we have not used any experimental data to obtain our final results.
\begin{table}[h!]
\begin{center}
\scalebox{0.8}{
\renewcommand{\arraystretch}{1.8}
\begin{tabular}{c|c|c|c|c|c|c|c|c|c|c}
$c_s$  & $c_p$  & $\sigma$ (fb) & $\mathcal A_1$ & $\mathcal A_2$ & $\mathcal A_3$ & $\mathcal A_4$ & $\mathcal A_5$ & $\mathcal A_6$ & $\mathcal A_7$ & error (1$\sigma$) \\
\hline\hline 
 0.8   &  0.8   &   3.11        &     2.98       &   2.65         &   -2.72        &   -0.98        &   -1.91        &    -0.53       &   -0.03        &   0.05 \\
 0.8   &  1.0   &   3.58        &     3.21       &   2.68         &   -2.86        &   -1.01        &   -1.98        &    -0.45       &   -0.01        &   0.05 \\
 0.9   &  0.8   &   3.71        &     2.84       &   2.57         &   -2.51        &   -0.90        &   -1.77        &    -0.46       &    0.00        &   0.05 \\
 0.9   &  1.0   &   4.19        &     2.98       &   2.57         &   -2.71        &   -0.95        &   -1.87        &    -0.46       &    0.04        &   0.05 \\
 1.0   &  0.0   &   6.38        &     0.04       &  -0.06         &    0.02        &   -0.10        &    0.05        &     0.13       &    0.02        &   0.06 \\
 1.0   &  0.2   &   6.48        &     0.75       &   0.65         &   -0.73        &   -0.21        &   -0.53        &    -0.07       &    0.02        &   0.06 \\
 1.0   &  0.4   &   6.78        &     1.54       &   1.26         &   -1.33        &   -0.36        &   -0.86        &    -0.21       &    0.02        &   0.06 \\
 1.0   &  0.6   &   7.28        &     2.02       &   1.71         &   -1.89        &   -0.63        &   -1.20        &    -0.31       &    0.04        &   0.06 \\
 1.0   &  0.8   &   7.98        &     2.45       &   2.16         &   -2.29        &   -0.78        &   -1.43        &    -0.41       &    0.01        &   0.06 \\
 1.0   &  1.0   &   8.89        &     2.78       &   2.34         &   -2.54        &   -0.90        &   -1.73        &    -0.39       &    0.08        &   0.06 \\
\hline\hline
\end{tabular}}
\caption{The measured integrated asymmetries $\mathcal A_{1-7}$ (in $\%$) at NLO+PS for the set of observables $\mathcal O_{1-7}$ at LHC with
$\sqrt{S}$ = 13 TeV for the process $pp \to t\bar{t}H$ with dileptonic tops for various values of coupling $c_s$ and $c_p$ for $10^{7}$ 
events.}
\label{Asym_13TeV_NLO_PS}
\end{center}
\end{table}
\begin{table}[h!]
\begin{center}
\scalebox{0.8}{
\renewcommand{\arraystretch}{1.8}
\begin{tabular}{c|c|c|c|c|c|c|c|c|c|c}
$c_s$ &  $c_p$ & $\sigma$ (fb) & $\mathcal A_1$ & $\mathcal A_2$ & $\mathcal A_3$ & $\mathcal A_4$ & $\mathcal A_5$ & $\mathcal A_6$ & $\mathcal A_7$ & error (1$\sigma$) \\
\hline\hline
 0.8  &   0.8  &     9.70      &    2.93        &    2.54        &    -2.66       &    -0.93       &     -1.88      &    -0.49       &  -0.03         &  0.05  \\
 0.8  &   1.0  &    11.26      &    3.17        &    2.74        &    -2.91       &    -1.00       &     -2.01      &    -0.52       &  -0.02         &  0.05  \\
 0.9  &   0.8  &    11.54      &    2.78        &    2.43        &    -2.53       &    -0.86       &     -1.78      &    -0.46       &  -0.06         &  0.05 \\
 0.9  &   1.0  &    13.10      &    3.04        &    2.64        &    -2.76       &    -1.00       &     -1.92      &    -0.55       &   0.00         &  0.05 \\
 1.0  &   0.0  &    10.80      &   -0.09        &   -0.07        &     0.06       &     0.01       &     -0.04      &    -0.06       &  -0.04         &  0.05 \\
 1.0  &   0.2  &    10.98      &    0.82        &    0.76        &    -0.79       &    -0.38       &     -0.65      &    -0.16       &   0.00         &  0.05 \\
 1.0  &   0.4  &    11.52      &    1.51        &    1.33        &    -1.41       &    -0.51       &     -0.97      &    -0.23       &  -0.02         &  0.05 \\
 1.0  &   0.6  &    12.39      &    2.22        &    1.89        &    -1.94       &    -0.68       &     -1.30      &    -0.31       &   0.04         &  0.05 \\
 1.0  &   0.8  &    13.60      &    2.67        &    2.37        &    -2.43       &    -0.80       &     -1.63      &    -0.48       &   0.02         &  0.05 \\
 1.0  &   1.0  &    15.16      &    2.90        &    2.50        &    -2.63       &    -0.90       &     -1.80      &    -0.53       &   0.01         &  0.05 \\  
\hline\hline
\end{tabular}}
\caption{The measured integrated asymmetries $\mathcal A_{1-7}$ (in $\%$) at NLO+PS for the set of observables $\mathcal O_{1-7}$ at HL-LHC 
with $\sqrt{S}$ = 14 TeV for the process $pp \to t\bar{t}H$ with dileptonic tops for various values of $c_s$ and $c_p$ for $10^7$ 
events.}
\label{Asym_14TeV_NLO_PS}
\end{center}
\end{table}
\begin{table}[h!]
\begin{center}
\scalebox{0.8}{
\renewcommand{\arraystretch}{1.8}
\begin{tabular}{c|c|c|c|c|c|c|c|c|c|c}
$c_s$ &  $c_p$  & $\sigma$ (fb) &  $\mathcal A_1$ & $\mathcal A_2$ & $\mathcal A_3$ & $\mathcal A_4$ & $\mathcal A_5$ & $\mathcal A_6$ & $\mathcal A_7$ & error (1$\sigma$) \\
\hline\hline
 0.8  &   0.8   &   343.25      &      2.74       &      2.42      &     -2.85      &     -0.88      &     -1.76      &    -0.37       &     -0.04      &  0.07 \\
 0.8  &   1.0   &   501.84      &      2.91       &      2.77      &     -2.87      &     -0.86      &     -1.89      &    -0.61       &      0.03      &  0.06 \\
 0.9  &   0.8   &   495.49      &      2.60       &      2.56      &     -2.68      &     -0.78      &     -1.74      &    -0.48       &      0.01      &  0.06 \\
 0.9  &   1.0   &   576.18      &      2.85       &      2.64      &     -2.87      &     -0.83      &     -1.88      &    -0.42       &     -0.09      &  0.06 \\
 1.0  &   0.0   &   358.09      &     -0.01       &      0.07      &      0.05      &      0.00      &      0.04      &    -0.03       &     -0.06      &  0.07 \\
 1.0  &   0.2   &   365.24      &      0.98       &      0.77      &     -0.95      &     -0.29      &     -0.67      &    -0.10       &      0.11      &  0.07 \\
 1.0  &   0.4   &   386.94      &      1.64       &      1.47      &     -1.59      &     -0.36      &     -0.89      &    -0.28       &     -0.02      &  0.07 \\
 1.0  &   0.6   &   422.13      &      2.09       &      1.87      &     -2.19      &     -0.69      &     -1.51      &    -0.33       &      0.01      &  0.07 \\
 1.0  &   0.8   &   472.52      &      2.45       &      2.36      &     -2.64      &     -0.73      &     -1.61      &    -0.50       &     -0.09      &  0.07 \\
 1.0  &   1.0   &   535.98      &      2.87       &      2.48      &     -2.96      &     -0.91      &     -1.72      &    -0.41       &      0.03      &  0.07 \\
\hline\hline
\end{tabular}}
\caption{The measured integrated asymmetries $\mathcal A_{1-7}$ (in $\%$) at NLO+PS for the set of observables $\mathcal O_{1-7}$ at FCC-hh with
$\sqrt{S}$ = 100 TeV for the process $pp \to t\bar{t}H$ with dileptonic tops for various values of $c_s$ and $c_p$ for $10^7$ events.}
\label{Asym_100TeV_NLO_PS}
\end{center}
\end{table}

In order to arrive at the functional form of various asymmetries for different LHC energies, we fit the data obtained
for various values of $c_s$ and $c_p$ for a given $\mathcal CP$ asymmery at a time with the assumption that the cross-section is
consistent with the Standard Model cross-section for a given LHC energy. Based on the numerical results given in Table
\ref{Asym_13TeV_NLO_PS}, we expect the following functional form of cross-section and asymmetries $\mathcal A_1$, $\mathcal A_2$,
$\mathcal A_3$, $\mathcal A_4$, $\mathcal A_5$ and $\mathcal A_6$ at LHC for $\sqrt{S}$ = 13 TeV at NLO+PS accuracy:
\begin{eqnarray}
\label{func_Asy13_NLO}
\nonumber
\sigma^{13 TeV}_{NLO+PS}        &=& 6.38~c_s^2 + 2.51~c_p^2, \\
\nonumber
\mathcal A^{13 TeV}_{1, NLO+PS}  &=& \frac{3.87 - 0.26~c_p^2 + 24.82~c_p~c_s - 3.69~c_s^2}{\sigma^{13 TeV}_{NLO+PS}}, \\
\nonumber
\mathcal A^{13 TeV}_{2, NLO+PS}  &=& \frac{5.06 - 2.95~c_p^2 + 24.45~c_p~c_s - 5.61~c_s^2}{\sigma^{13 TeV}_{NLO+PS}}, \\
\nonumber
\mathcal A^{13 TeV}_{3, NLO+PS}  &=& \frac{-2.80 + 1.02~c_p^2 - 23.74~c_p~c_s + 2.91~c_s^2}{\sigma^{13 TeV}_{NLO+PS}}, \\
\nonumber
\end{eqnarray}
\begin{eqnarray}
\label{func_Asy13_NLO_dis}
\nonumber
\mathcal A^{13 TeV}_{4, NLO+PS}  &=& \frac{-1.04 - 1.82~c_p^2 - 5.70~c_p~c_s + 0.68~c_s^2}{\sigma^{13 TeV}_{NLO+PS}}, \\
\nonumber
\mathcal A^{13 TeV}_{5, NLO+PS}  &=& \frac{-3.66 + 0.22~c_p^2 - 15.34~c_p~c_s + 3.77~c_s^2}{\sigma^{13 TeV}_{NLO+PS}}, \\
\mathcal A^{13 TeV}_{6, NLO+PS}  &=& \frac{-1.87 + 2.59~c_p^2 - 7.02~c_p~c_s + 2.74~c_s^2}{\sigma^{13 TeV}_{NLO+PS}}.
\end{eqnarray}

Similarly, the functional form obtained for HL-LHC with $\sqrt{S}$ = 14 TeV and FCC-hh with $\sqrt{S}$ = 100 TeV at NLO+PS from the numerical results
given in Tables \ref{Asym_14TeV_NLO_PS} and \ref{Asym_100TeV_NLO_PS} will be:
\begin{eqnarray}
\label{func_Asy14_NLO}
\nonumber
\sigma^{14 TeV}_{NLO+PS}        &=& 10.81~c_s^2 + 4.34~c_p^2, \\
\nonumber
\mathcal A^{14 TeV}_{1, NLO+PS}  &=& \frac{0.34 - 2.63~c_p^2 + 47.95~c_p~c_s - 1.18~c_s^2}{\sigma^{14 TeV}_{NLO+PS}}, \\
\nonumber
\mathcal A^{14 TeV}_{2, NLO+PS}  &=& \frac{0.33 - 2.69~c_p^2 + 41.82~c_p~c_s - 0.85~c_s^2}{\sigma^{14 TeV}_{NLO+PS}}, \\
\nonumber
\mathcal A^{14 TeV}_{3, NLO+PS}  &=& \frac{-0.78 + 1.35~c_p^2 - 41.94~c_p~c_s + 1.06~c_s^2}{\sigma^{14 TeV}_{NLO+PS}}, \\
\nonumber
\mathcal A^{14 TeV}_{4, NLO+PS}  &=& \frac{-0.91 + 0.54~c_p^2 - 13.87~c_p~c_s + 0.46~c_s^2}{\sigma^{14 TeV}_{NLO+PS}}, \\
\nonumber
\mathcal A^{14 TeV}_{5, NLO+PS}  &=& \frac{-2.51 - 0.07~c_p^2 - 26.37~c_p~c_s + 1.63~c_s^2}{\sigma^{14 TeV}_{NLO+PS}}, \\
\mathcal A^{14 TeV}_{6, NLO+PS}  &=& \frac{1.63 - 2.82~c_p^2 - 4.54~c_p~c_s - 2.23~c_s^2}{\sigma^{14 TeV}_{NLO+PS}}.
\end{eqnarray}
\begin{eqnarray}
\label{func_Asy100_NLO}
\nonumber
\sigma^{100 TeV}_{NLO+PS}        &=&  253.26~c_p^2 + 343.16~c_s^2, \\
\nonumber
\mathcal A^{100 TeV}_{1, NLO+PS}  &=& \frac{-44.98 + 287.24~cp^2 + 1348.11~cp~cs + 68.75~cs^2}{\sigma^{100 TeV}_{NLO+PS}}, \\
\nonumber
\mathcal A^{100 TeV}_{2, NLO+PS}  &=& \frac{-36.14 + 355.57~cp^2 + 1147.06~cp~cs + 64.27~cs^2}{\sigma^{100 TeV}_{NLO+PS}}, \\
\nonumber
\mathcal A^{100 TeV}_{3, NLO+PS}  &=& \frac{89.60 - 209.46~cp^2 - 1506.58~cp~cs - 83.36~cs^2}{\sigma^{100 TeV}_{NLO+PS}}, \\
\nonumber
\mathcal A^{100 TeV}_{4, NLO+PS}  &=& \frac{10.76 - 112.51~cp^2 - 388.40~cp~cs - 12.76~cs^2}{\sigma^{100 TeV}_{NLO+PS}}, \\
\nonumber
\mathcal A^{100 TeV}_{5, NLO+PS}  &=& \frac{-28.57 - 120.45~cp^2 - 938.67~cp~cs + 27.26~cs^2}{\sigma^{100 TeV}_{NLO+PS}}, \\
\mathcal A^{100 TeV}_{6, NLO+PS}  &=& \frac{33.16 - 107.71~cp^2 - 170.47~cp~cs - 41.57~cs^2}{\sigma^{100 TeV}_{NLO+PS}}.
\end{eqnarray}

From the expressions \ref{func_Asy13_NLO}, \ref{func_Asy14_NLO} and \ref{func_Asy100_NLO}, we notice that the 
asymmetry arises mostly from the interfering term i.e. the term proportional to $c_s~c_p$ while the constant term 
and the term proportional to $c_s^2$ and $c_p^2$ contribute less.

\begin{figure}[]
\centering
\includegraphics[width=7.3cm,height=7.5cm]{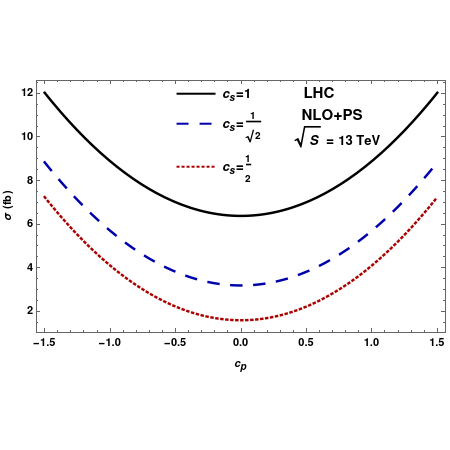}
\includegraphics[width=7.3cm,height=7.5cm]{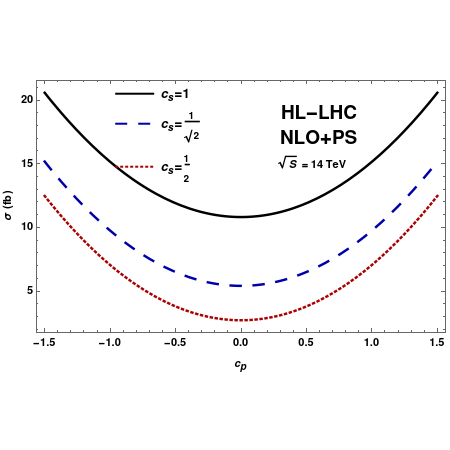}
\includegraphics[width=7.3cm,height=7.5cm]{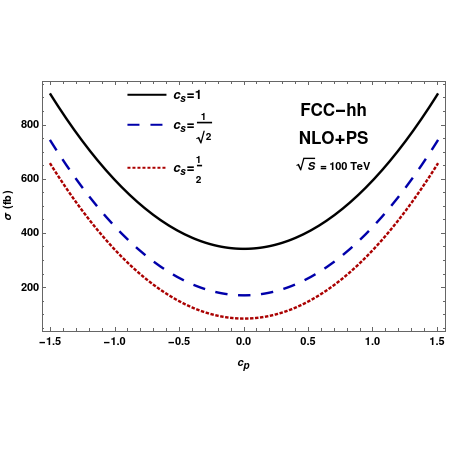}
\caption{\small Cross-section as a function of anomalous coupling $c_p$ at three different values of coupling $c_s$, viz, $c_s$ = 1,
$\frac{1}{\sqrt{2}}$, and $\frac{1}{2}$ at LHC with $\sqrt{S}$ = 13 TeV (upper left), HL-LHC with $\sqrt{S}$ = 14 TeV (upper right) and
FCC-hh with $\sqrt{S}$ = 100 TeV (lower) at NLO+PS accuracy.}
\label{xs_plts_NLO_PS}
\end{figure}

In Fig. \ref{xs_plts_NLO_PS}, we show the cross-section calculated at NLO+PS as a function of pseudoscalar coupling $c_p$ for three 
different parameter points: $c_s$ = $1, \frac{1}{\sqrt{2}}$ and $\frac{1}{2}$ at LHC with $\sqrt{S}$ = 13 TeV for an integrated 
luminosity of 139 fb$^{-1}$, HL-LHC with $\sqrt{S}$ = 14 TeV for the projected luminosity of 3 ab$^{-1}$ and FCC-hh with $\sqrt{S}$ = 
100 TeV for the projected luminosity of 30 ab$^{-1}$. We see that the cross-section is symmetric around $c_p$ = 0 and is equally 
sensitive to both positive and negative values of the pseudoscalar coupling $c_p$. Furthermore, we observe that the presence of 
$\mathcal CP$-violating coupling $c_p$ in $t\bar{t}H$ interaction increases the cross-section. Particularly, the variation of $c_p$ at 
a given value of $c_s$ leads to a significant increase in the value of cross-section. For all the parameter 
points, the cross-section can increase as much as two times or more by varying the pseudo-scalar coupling $c_p$. However, the 
cross-section gets strongest contribution corresponding to the point $c_s$ = 1. This indicates that the cross-section is sensitive to 
the $\mathcal CP$-violating part of the $t\bar{t}H$ interaction and can be used to probe pseudoscalar coupling $c_p$.
Individual bounds on pseudoscalar coupling $c_p$ at NLO+PS at three different values of $c_s$, viz, $c_s$ = 1, $\frac{1}{2}$ and $\frac{1}{\sqrt{2}}$ obtained from the 
cross-section measurements at LHC with $\sqrt{S}$ = 13 TeV, HL-LHC with $\sqrt{S}$ = 14 TeV and FCC-hh with $\sqrt{S}$ = 100 TeV are shown in Table \ref{xsec_bounds_NLO_PS}.
\begin{table}[h!]
\begin{center}
\scalebox{1.0}{
\renewcommand{\arraystretch}{2.2}
\begin{tabular}{c|c|c|c}
$c_s$                                                       &  1                          &  $\frac{1}{\sqrt{2}}$      &  $\frac{1}{2}$ \\
\hline\hline
$\frac{\Delta \sigma}{\sigma}_{pp \to t\bar{t}H}^{LHC}$     &  $-0.16 \le c_p \le 0.16$   &  $-1.14 \le c_p \le 1.14$  &  $-1.39\le c_p \le 1.39$ \\
\hline
$\frac{\Delta \sigma}{\sigma}_{pp \to t\bar{t}H}^{HL-LHC}$  &  $-0.81 \le c_p \le 0.81$   &  $-1.38 \le c_p \le 1.38$  &  $-1.59 \le c_p \le 1.59$ \\
\hline
$\frac{\Delta \sigma}{\sigma}_{pp \to t\bar{t}H}^{FCC-hh}$  &  $-0.54 \le c_p \le 0.54$   &  $-0.98 \le c_p \le 0.98$  &  $-1.14 \le c_p \le 1.14$ \\
\hline\hline
\end{tabular}}
\caption{Individual constraints on anomalous coupling $c_p$ at three different values of $c_s$, viz, $c_s$ = 1, $\frac{1}{\sqrt{2}}$ and
$\frac{1}{2}$ at 2.5$\sigma$ C.L. obtained from $t\bar{t}H$ production cross-section at the LHC with $\sqrt{S}$ = 13 TeV, HL-LHC
with $\sqrt{S}$ = 14 TeV and FCC-hh with $\sqrt{S}$ = 100 TeV respectively at NLO+PS accuracy.}
\label{xsec_bounds_NLO_PS}
\end{center}
\end{table}

\begin{figure}[]
\centering
\includegraphics[width=8.0cm,height=7.3cm]{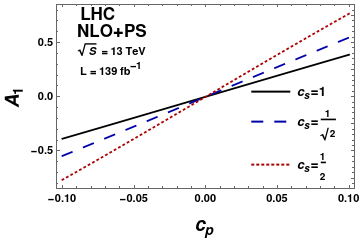}
\includegraphics[width=8.0cm,height=7.3cm]{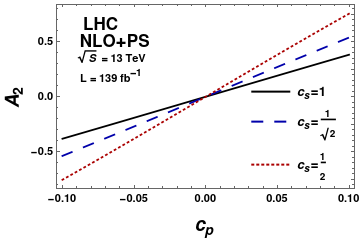}
\includegraphics[width=8.0cm,height=7.3cm]{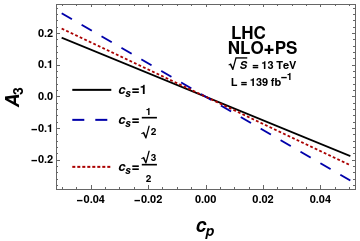}
\includegraphics[width=8.0cm,height=7.3cm]{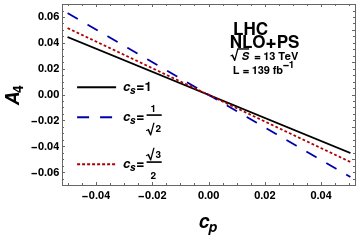}
\includegraphics[width=8.0cm,height=7.3cm]{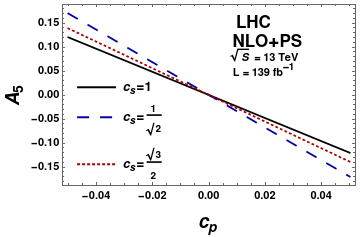}
\includegraphics[width=8.0cm,height=7.3cm]{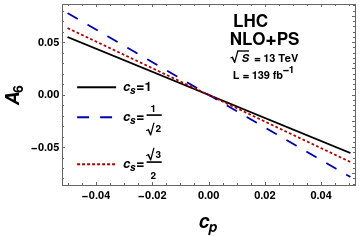}
\caption{\small Asymmetry as a function of anomalous coupling $c_p$ at three different values of coupling $c_s$, viz, $c_s$ = 1, $\frac{1}{\sqrt{2}}$ and $\frac{1}{2}$ at LHC 
with $\sqrt{S}$ = 13 TeV corresponding to the observables $\mathcal O_1$, $\mathcal O_2$, $\mathcal O_3$, $\mathcal O_4$, $\mathcal O_5$, and $\mathcal O_6$ at NLO+PS .}
\label{Asy13_plots_NLO_PS}
\end{figure}

\begin{figure}[]
\centering
\includegraphics[width=8.0cm,height=7.3cm]{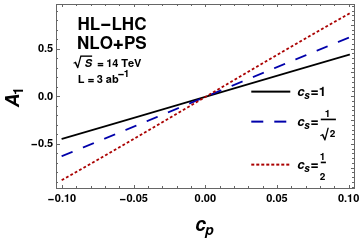}
\includegraphics[width=8.0cm,height=7.3cm]{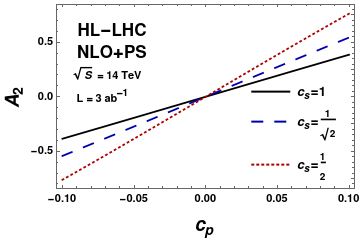}
\includegraphics[width=8.0cm,height=7.3cm]{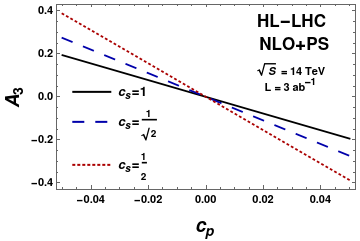}
\includegraphics[width=8.0cm,height=7.3cm]{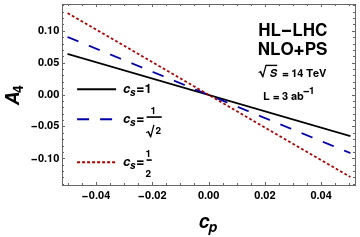}
\includegraphics[width=8.0cm,height=7.3cm]{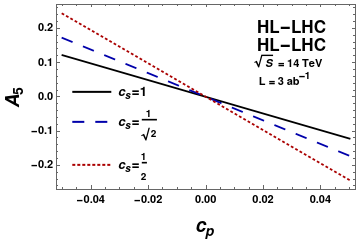}
\includegraphics[width=8.0cm,height=7.3cm]{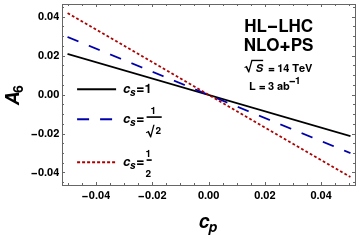}
\caption{\small Asymmetry as a function of anomalous coupling $c_p$ at three different values of coupling $c_s$, viz, $c_s$ = 1, $\frac{1}{\sqrt{2}}$ and $\frac{1}{2}$ at 
HL-LHC with $\sqrt{S}$ = 14 TeV corresponding to the observables $\mathcal O_1$, $\mathcal O_2$, $\mathcal O_3$, $\mathcal O_4$, $\mathcal O_5$, and $\mathcal O_6$ at 
NLO+PS.}
\label{Asy14_plots_NLO_PS}
\end{figure}

\begin{figure}[]
\centering
\includegraphics[width=8.0cm,height=7.3cm]{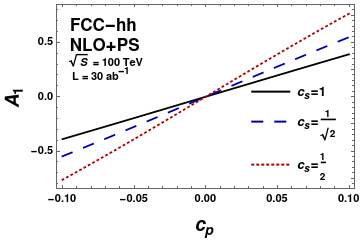}
\includegraphics[width=8.0cm,height=7.3cm]{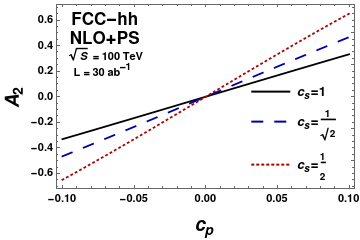}
\includegraphics[width=8.0cm,height=7.3cm]{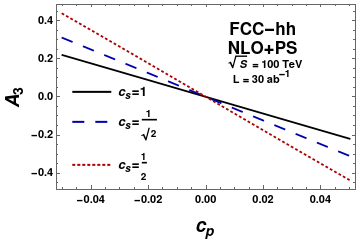}
\includegraphics[width=8.0cm,height=7.3cm]{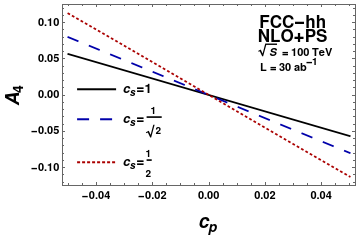}
\includegraphics[width=8.0cm,height=7.3cm]{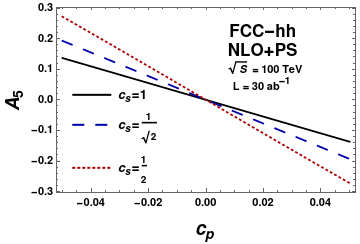}
\includegraphics[width=8.0cm,height=7.3cm]{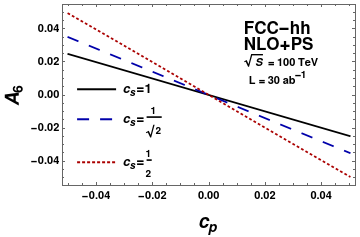}
\caption{\small Asymmetry as a function of anomalous coupling $c_p$ at three different values of coupling $c_s$, viz, $c_s$ = 1, $\frac{1}{\sqrt{2}}$ and $\frac{1}{2}$ at 
FCC-hh with $\sqrt{S}$ = 100 TeV corresponding to the observables $\mathcal O_1$, $\mathcal O_2$, $\mathcal O_3$, $\mathcal O_4$, $\mathcal O_5$, and $\mathcal O_6$ at 
NLO+PS.}
\label{Asy100_plots_NLO_PS}
\end{figure}

In Figs. \ref{Asy13_plots_NLO_PS}, \ref{Asy14_plots_NLO_PS} and \ref{Asy100_plots_NLO_PS}, we show the production asymmetry calculated 
at NLO+PS corresponding to the observables $\mathcal O_1$, $\mathcal O_2$, $\mathcal O_3$, $\mathcal O_4$, $\mathcal O_5$ and $\mathcal 
O_6$ as a function of pseudoscalar coupling $c_p$ at different values of scalar coupling $c_s$ at LHC with $\sqrt{S}$ = 13 TeV for an 
integrated luminosity of 139 fb$^{-1}$, HL-LHC with $\sqrt{S}$ = 14 TeV for the projected luminosity of 3 ab$^{-1}$ and FCC-hh with 
$\sqrt{S}$ = 100 TeV for the projected luminosity of 30 ab$^{-1}$, respectively. Three illustrative values of scalar coupling, 
$c_s~=~1,~\frac{1}{\sqrt{2}},~\rm{and}~\frac{1}{2}$ are considered. We notice that the asymmetry is highly sensitive to the 
pseudoscalar coupling $c_p$ which can be seen from the Figs. \ref{Asy13_plots_NLO_PS}, \ref{Asy14_plots_NLO_PS} and 
\ref{Asy100_plots_NLO_PS}, as the value of coupling $c_p$ changes for a given value of coupling $c_s$, the asymmetry changes 
significantly. Furthermore, we notice that the asymmetry is symmetric around $c_p$ = 0 and is equally sensitive to both positive and 
negative values of coupling $c_p$. Interestingly, the asymmetry increases by changing the coupling $c_p$.

\begin{figure}[]
\centering
\includegraphics[width=8.0cm,height=7.3cm]{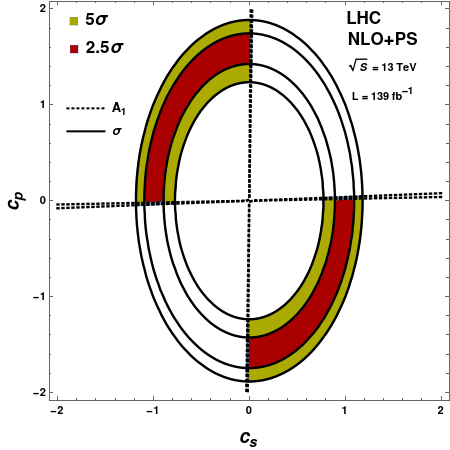}
\includegraphics[width=8.0cm,height=7.3cm]{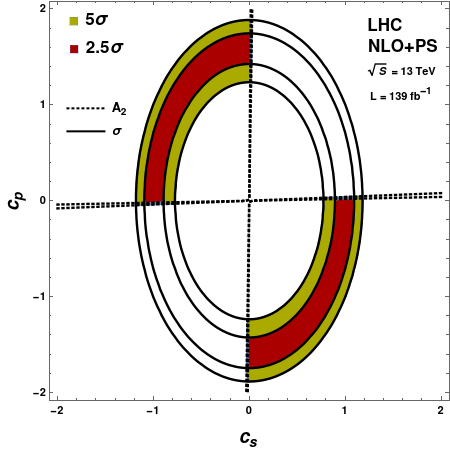}
\includegraphics[width=8.0cm,height=7.3cm]{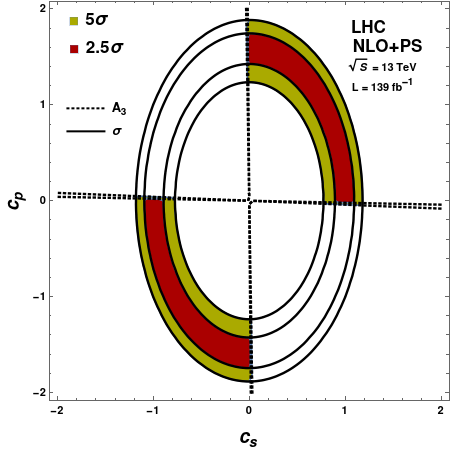}
\includegraphics[width=8.0cm,height=7.3cm]{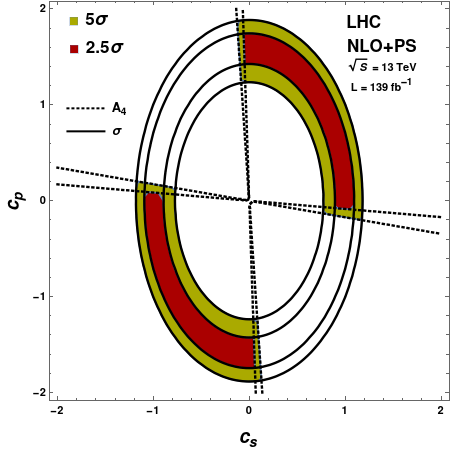}
\includegraphics[width=8.0cm,height=7.3cm]{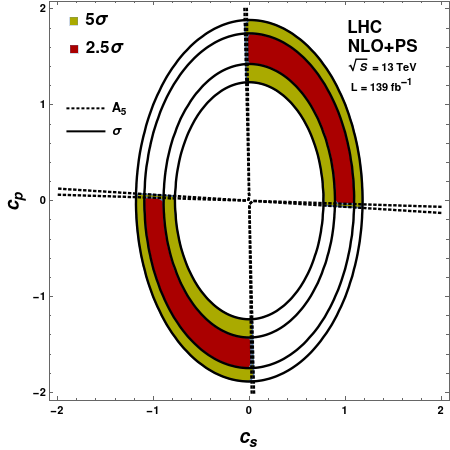}
\includegraphics[width=8.0cm,height=7.3cm]{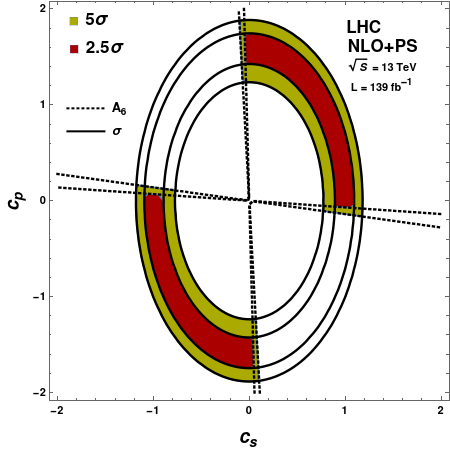}
\caption{\small Contour plots of cross-section and production asymmetry measured at NLO+PS accuracy in $c_p-c_s$ plane at LHC with $\sqrt{S}$ = 13 TeV for the observables 
$\mathcal O_1$, $\mathcal O_2$ (top row), $\mathcal O_3$, $\mathcal O_4$ (middle row), $\mathcal O_5$ and $\mathcal O_6$ (lower row). The red and yellow area represent the 
parameter space allowed at 2.5$\sigma$ and 5$\sigma$ C.L. respectively from combined measurements of cross-section and production asymmetry.}
\label{contplts13TeV_NLO_PS}
\end{figure}

\begin{figure}[]
\centering
\includegraphics[width=8.0cm,height=7.3cm]{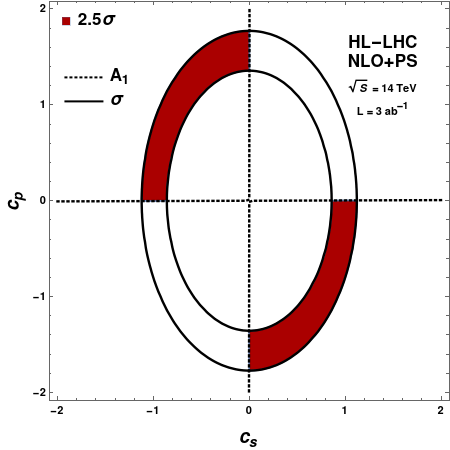}
\includegraphics[width=8.0cm,height=7.3cm]{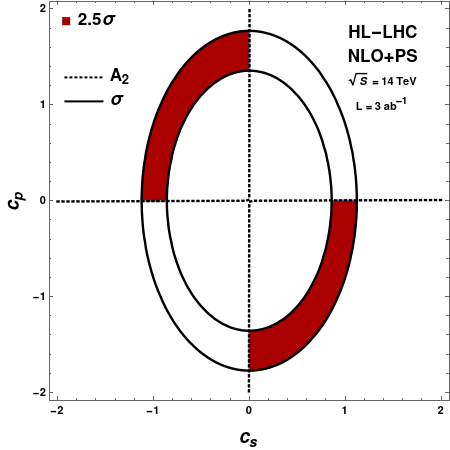}
\includegraphics[width=8.0cm,height=7.3cm]{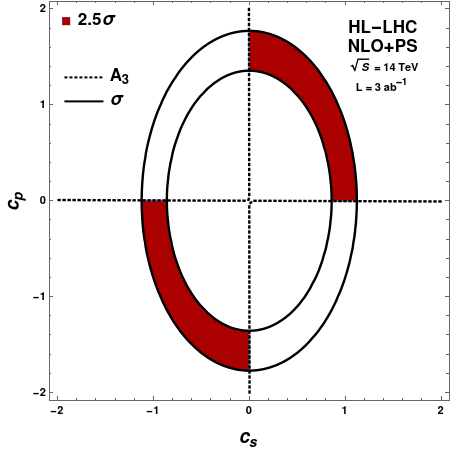}
\includegraphics[width=8.0cm,height=7.3cm]{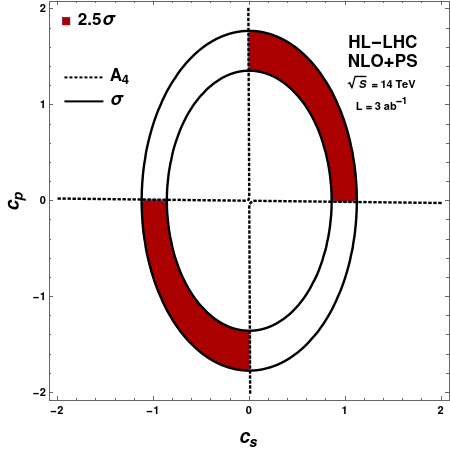}
\includegraphics[width=8.0cm,height=7.3cm]{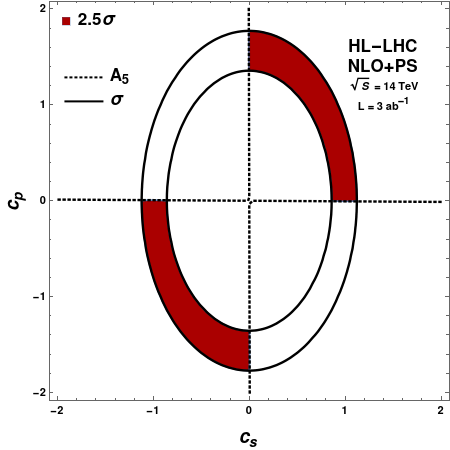}
\includegraphics[width=8.0cm,height=7.3cm]{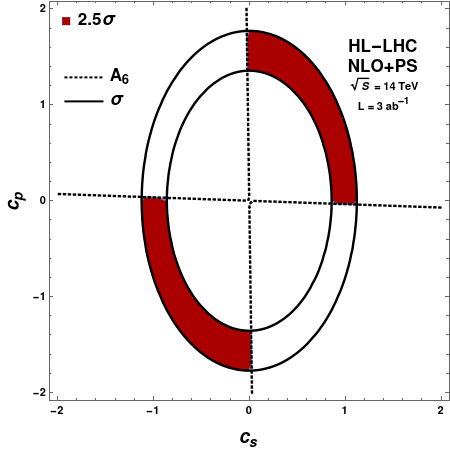}
\caption{\small Contour plots of cross-section and production asymmetry measured at NLO+PS accuracy in $c_p-c_s$ plane at HL-LHC with $\sqrt{S}$ = 14 TeV for the observables 
$\mathcal O_1$, $\mathcal O_2$ (top row), $\mathcal O_3$,$\mathcal O_4$ (middle row), $\mathcal O_5$ and $\mathcal O_6$ (lower row). The red and yellow area represent the 
parameter space allowed at 2.5$\sigma$ and 5$\sigma$ C.L. respectively from combined measurements of cross-section and production asymmetry.}
\label{contplts14TeV_NLO_PS}
\end{figure}

\begin{figure}[]
\centering
\includegraphics[width=8.0cm,height=7.3cm]{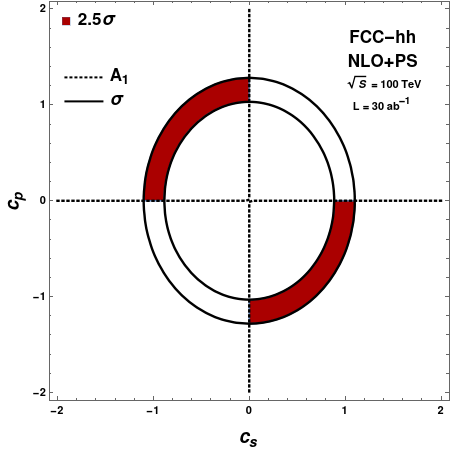}
\includegraphics[width=8.0cm,height=7.3cm]{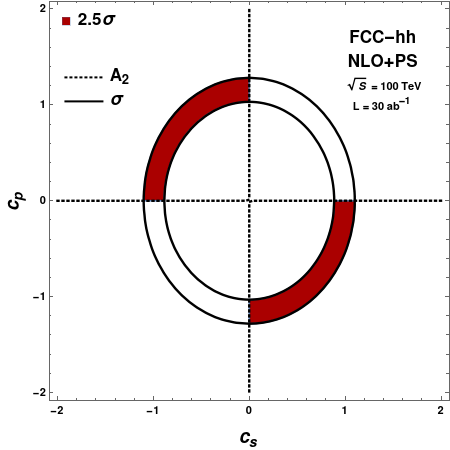}
\includegraphics[width=8.0cm,height=7.3cm]{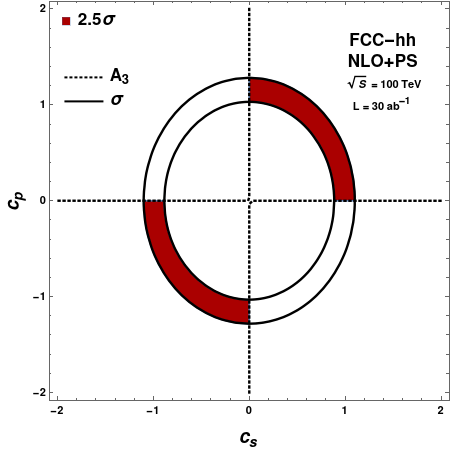}
\includegraphics[width=8.0cm,height=7.3cm]{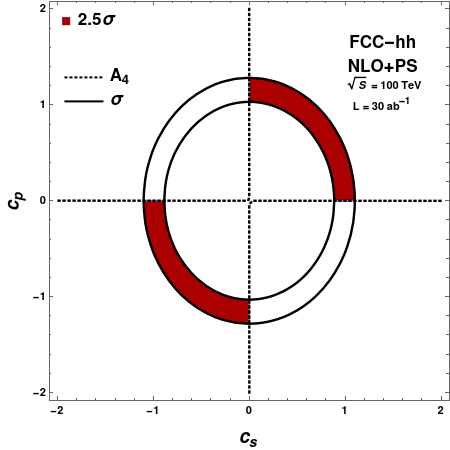}
\includegraphics[width=8.0cm,height=7.3cm]{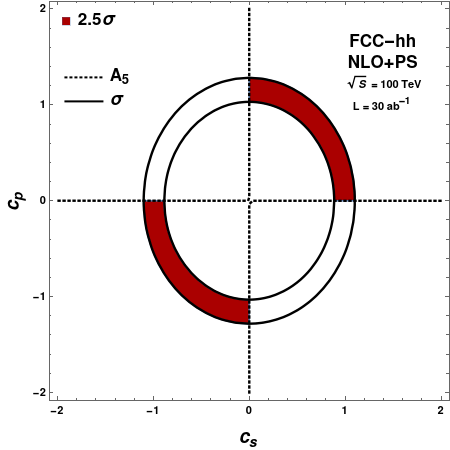}
\includegraphics[width=8.0cm,height=7.3cm]{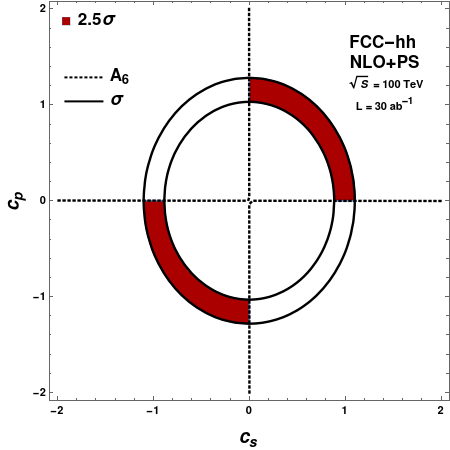}
\caption{\small Contour plots of cross-section and production asymmetry measured at NLO+PS accuracy in $c_p-c_s$ plane at FCC-hh with $\sqrt{S}$ = 100 TeV for the observables 
$\mathcal O_1$, $\mathcal O_2$ (top row), $\mathcal O_3$,$\mathcal O_4$ (middle row), $\mathcal O_5$ and $\mathcal O_6$ (lower row). The red and yellow area represent the 
parameter space allowed at 2.5$\sigma$ and 5$\sigma$ C.L. respectively from combined measurements of cross-section and production asymmetry.}
\label{contplts100TeV_NLO_PS}
\end{figure}

In Figs. \ref{contplts13TeV_NLO_PS}, we show 2.5$\sigma$ and 5$\sigma$ regions in $c_p-c_s$ plane allowed by the combined measurements 
of cross-section and production asymmetry measured at NLO+PS for the observables $\mathcal O_1$, $\mathcal O_2$, $\mathcal O_3$, 
$\mathcal O_4$, $\mathcal O_5$ and $\mathcal O_6$ for the process $pp \to t\bar{t}H$ in the dileptonic decay channel of top-quark at 
the LHC with $\sqrt{S}$ = 13 TeV for an integrated luminosity of 139 fb$^{-1}$ and 2.5$\sigma$ regions for HL-LHC with $\sqrt{S}$ = 14 
TeV and FCC-hh with $\sqrt{S}$ = 100 TeV for the projected luminosities of 3 ab$^{-1}$ and 30 ab$^{-1}$ in Figs. 
\ref{contplts14TeV_NLO_PS} and \ref{contplts100TeV_NLO_PS} respectively. The solid lines represent the region allowed by the 
cross-section and the dotted lines represent the region allowed by the production asymmetry. The complementarity of both measurements 
is beautifully depicted here: the intersection of the circle shaped region from cross-section and the lines from production asymmetry 
gives much more stringent bounds than the separate measurements. The combination of both measurements is very powerful and almost 
removes the large $c_p$ region present in the limit from production asymmetry. The resulting combined regions are shown in red and 
yellow which corresponds to the region allowed at 2.5$\sigma$ and 5$\sigma$ C.L. respectively. From here we can give a rough estimate 
of the constraint on the coupling $c_p$ from the combined measurement of $\sigma$ and production asymmetry. However, we conduct a more 
detailed analysis regarding the sensitivity of the observables $\mathcal O_1$, $\mathcal O_2$, $\mathcal O_3$, $\mathcal O_4$, 
$\mathcal O_5$ and $\mathcal O_6$ to the $\mathcal CP$-violating Higgs-top coupling $c_p$ for three different parameter points: $c_s$ 
= 1, $\frac{1}{\sqrt{2}}$ and $\frac{1}{2}$ at LHC with $\sqrt{S}$ = 13 TeV and an integrated luminosity of 139 fb$^{-1}$ and for the 
Future Hadron Colliders, namely HL-LHC and FCC-hh with ($\sqrt{S},~\int L dt$) = (14 TeV, 3 ab$^{-1}$) and (100 TeV, 30 ab$^{-1}$) 
respectively. The results corresponding to these measurements at 2.5$\sigma$ C.L. are presented in Table \ref{Asym_bounds_NLO_PS}.
\begin{table}[h!]
\begin{center}
\scalebox{1.0}{
\renewcommand{\arraystretch}{1.8}
\begin{tabular}{c|c|c|c|c|c|c}
Collider  &  $\sqrt{S}$, $\mathcal \int Ldt$  &  Asymmetry       & \multicolumn{4}{c}{$ c_p \times(10^{-2})$} \\
\hline
          &                                   &                  & $c_s$ &  1                       &  $\frac{1}{\sqrt{2}}$        & $\frac{1}{2}$  \\
\hline\hline
LHC       &  13 TeV, 139 fb$^{-1}$            & $\mathcal A_1$   &       &  $c_p \le 1.96 $         &  $c_p \le 1.38 $             &  $c_p \le 0.98 $, \\
          &                                   & $\mathcal A_2$   &       &  $c_p \le 1.99 $         &  $c_p \le 1.41 $             &  $c_p \le 0.99 $, \\
          &                                   & $\mathcal A_3$   &       &  $-2.05 \le c_p $        & $-1.45 \le c_p $             & $-1.02 \le c_p $, \\
          &                                   & $\mathcal A_4$   &       &  $-8.55 \le c_p $        & $-6.04 \le c_p $             & $-4.27 \le c_p $, \\
          &                                   & $\mathcal A_5$   &       &  $-3.17 \le c_p $        & $-2.24 \le c_p $             & $-1.58 \le c_p $, \\
          &                                   & $\mathcal A_6$   &       &  $-6.93 \le c_p $        & $-4.90 \le c_p $             & $-3.47 \le c_p $, \\
\hline
HL-LHC    &  14 TeV, 3 ab$^{-1}$              & $\mathcal A_1$   &       &  $c_p \le 0.34 $         & 	 $c_p \le 0.24 $             &  $c_p \le 0.17 $, \\
          &                                   & $\mathcal A_2$   &       &  $c_p \le 0.38 $         &  $c_p \le 0.27 $             &  $c_p \le 0.19 $, \\
          &                                   & $\mathcal A_3$   &       &  $-0.38 \le c_p $        & $-0.27 \le c_p $             &  $-0.19 \le c_p $, \\
          &                                   & $\mathcal A_4$   &       &  $-1.16 \le c_p $        & $-0.82 \le c_p $             &  $-0.58 \le c_p $, \\
          &                                   & $\mathcal A_5$   &       &  $-0.61 \le c_p $        & $-0.43 \le c_p $             &  $-0.30 \le c_p $, \\
          &                                   & $\mathcal A_6$   &       &  $-3.54 \le c_p $        & $-2.51 \le c_p $             &  $-1.77 \le c_p $, \\
\hline
FCC-hh    &  100 TeV, 30 ab$^{-1}$            & $\mathcal A_1$   &       & $c_p \le 0.016 $         &  $c_p \le 0.011 $            & $c_p \le 0.008 $, \\
          &                                   & $\mathcal A_2$   &       & $c_p \le 0.019 $         &  $c_p \le 0.013 $            & $c_p \le 0.009 $, \\
          &                                   & $\mathcal A_3$   &       & $-0.014 \le c_p $        & $-0.010 \le c_p $            & $-0.007 \le c_p $, \\
          &                                   & $\mathcal A_4$   &       & $-0.055 \le c_p $        & $-0.039 \le c_p $            & $-0.028 \le c_p $, \\
          &                                   & $\mathcal A_5$   &       & $-0.023 \le c_p $        & $-0.016 \le c_p $            & $-0.011 \le c_p $, \\
          &                                   & $\mathcal A_6$   &       & $-0.125 \le c_p $        & $-0.089 \le c_p $            & $-0.063 \le c_p $, \\
\hline\hline
\end{tabular}}
\caption{Individual constraints on anomalous coupling $c_p$ at NLO+PS accuracy at three different values of $c_s$, viz, $c_s$ = 1, $\frac{1}{2}$, $\frac{1}{\sqrt{2}}$ at 
2.5$\sigma$ C.L. allowed by the production asymmetries $\mathcal A_1$, $\mathcal A_2$, $\mathcal A_3$, $\mathcal A_4$, $\mathcal A_5$, and $\mathcal A_6$ corresponding to the 
observables $\mathcal O_1$, $\mathcal O_2$, $\mathcal O_3$, $\mathcal O_4$, $\mathcal O_5$ and $\mathcal O_6$ at LHC, HL-LHC and FCC-hh with $\sqrt{S}$ = 13 TeV, 14 TeV and 
100 TeV respectively and the luminosities of 139 fb$^{-1}$ to 30 ab $^{-1}$ have been explored.}
\label{Asym_bounds_NLO_PS}
\end{center}
\end{table}

Let us now compare our findings with the existing limits on pseudoscalar coupling $c_p$ derived from previous literature. In Ref. 
\cite{Bahl:2021dnc}, the authors applied machine-learning based interference to derive the expected constraints on $\mathcal 
CP$-violating top-Yukawa coupling in the context of top-associated Higgs production with the Higgs boson decaying to two photons and 
found that a $\mathcal CP$-odd top Yukawa coupling at the LHC can be constrained to $-0.8< c_p < 0.8$ and $-0.5 < c_p < 0.5$ at 
68.3$\%$ C.L. level for the luminosities of 139 fb$^{-1}$ and 300 fb$^{-1}$ respectively and for HL-LHC with a projected luminosity of 
3000 fb$^{-1}$, a bound of $-0.25 < c_p < 0.25$ can be found. These results were obtained at LO with a Higgs decay to two photons, 
whereas our findings are at NLO+PS with a stable Higgs. In Ref. \cite{Kobakhidze:2016mfx}, the $\mathcal CP$-odd 
component $c_p$ is constrained at 2$\sigma$ level to $c_p < 0.37$ by combining the LHC Run-1 and -2 Higgs data sets. However, EDM 
measurements placed the strongest constraints on the $\mathcal CP$-violating Higgs-top couplings and found an upper limit of 0.01 on 
the $c_p$ \cite{Brod:2013cka}.

\section{Results and Discussion} 
\label{R&D}

The discovery of the Higgs boson at the LHC furnishes new opportunities for exploring physics beyond the SM. Since its discovery at 
the Large Hadron Collider, it has always been crucial to investigate its $\mathcal CP$-properties as it is expected that it can 
provide an explanation of the fundamental question of matter–antimatter asymmetry. Probing the top-Yukawa coupling is important for 
measuring the $\mathcal CP$ nature of Higgs boson. In this study, we have explored the anomalous Higgs-top coupling and discussed the 
$\mathcal CP$-violation effects of Higgs-top interaction in the associated production of top-pair with Higgs boson, followed by the 
dileptonic decay of top-quark by means of T-odd observables constructed via the momenta of final state particles. Our 
results are obtained in a fully automatic way following the Higgs characterization model at next-to-leading order accuracy in QCD, 
including the parton shower effects. In particular, we derive constraints on the $\mathcal CP$-violating component of the $Ht\bar{t}$ 
interaction.

We measured the cross-section for the associated production of Higgs boson with top-pair at NLO+PS accuracy and show that the 
cross-section can be a crucial ingredient for finding the $\mathcal CP$-violation sensitivity to anomalous Higgs-top interaction. The 
measured cross-section was used to set constraints on the $\mathcal CP$-violating coupling $c_p$. The constraints were obtained at 
2.5$\sigma$ C.L. for LHC with $\sqrt{S}$ = 13 TeV for the integrated luminosity of 139 fb$^{-1}$ and for future hadron colliders, viz, 
HL-LHC and FCC-hh with $\sqrt{S}$ = 14 TeV and 100 TeV with projected luminosities of 3 ab$^{-1}$ and 30 ab$^{-1}$. The bounds on the 
pseudoscalar coupling $c_p$ obtained from cross-section measurements are presented in Table \ref{xsec_bounds_NLO_PS} for three 
parameter points: $c_s$ = 1, $\frac{1}{\sqrt{2}}$ and $\frac{1}{2}$ at 2.5$\sigma$ C.L. Ref. \cite{Bahl:2020wee} obtained constraints 
on the $\mathcal CP$-violating coupling $c_p$ in the interval [-0.3,0.3] at 1$\sigma$ level in the 2D parameterization, where the 
significant contribution originates from the production of Higgs boson through gluon-gluon annihilation and decay of Higgs boson to 
two photons. It has also been shown that allowing additional freedom to Higgs coupling weakens the constraints.

The production asymmetries were measured corresponding to the observables defined in Eq. \ref{observables} at NLO+PS and showed that 
the observables provide a sensitive probe for $\mathcal CP$-violation in the Higgs-top interaction. We find that the $\mathcal
CP$-violating component $c_p$ has been constrained corresponding to the largest asymmetry $\mathcal A_1$ to its maximum value to 
1.96$\times 10^{-2}$ at 2.5$\sigma$ CL for $c_s$ = 1 for the LHC with $\sqrt{S}$ = 13 TeV and an integrated luminosity of 139 
fb$^{-1}$. The corresponding limits for its luminosity intense variant HL-LHC and Future Circular Collider FCC-hh are estimated to be 
to 0.34$\times 10^{-2}$ and 0.016$\times 10^{-2}$ for the projected luminosities of 3.0 ab$^{-1}$ and 30 ab$^{-1}$ respectively at 
2.5$\sigma$ CL. The limits obtained corresponding to all the asymmetries $\mathcal A_{1-6}$ are presented in Table \ref{Asym_bounds_NLO_PS}. In 
Ref. \cite{Azevedo:2022jnd}, the bounds on the $\mathcal CP$-violating anomalous $t\bar{t}H$ interaction were obtained using the 
angular separation of leptons in the $t\bar{t}$ center-of-mass frame and set an exclusion limit on $c_p$: $c_p$ = [-0.698, 0.698] at 
95$\%$ CL for HL-LHC. This study was performed with the production of Higgs boson with a top pair where Higgs decays into a $b\bar{b}$ 
pair. Our study is however based on sensitivity measurement to $\mathcal CP$-violating anomalous coupling using $\mathcal CP$-odd 
observables in the $t\bar{t}H$ system. In addition, studies related to finding the Higgs-top $\mathcal CP$ phase 
were performed in Refs. \cite{Barman:2022pip,ATLAS:2023cbt}. To the best of our knowledge, this is the first study to measure the 
bounds on $\mathcal CP$-violating anomalous $t\bar{t}H$ coupling at NLO+PS order using $\mathcal CP$-odd observables constructed through 
momenta of the produced particles.

\section*{Acknowledgements}
This work was supported in part by University Grant Commission under a Start-Up Grant no. F30-377/2017 (BSR). We thank Surabhi Gupta for 
some valuable discussion. We acknowledge the use of computing facility at the general computing lab of Aligarh Muslim University.


\newpage

\begin{thebibliography}{58}

\bibitem{Novaes:1999yn}
S.~F.~Novaes,
[arXiv:hep-ph/0001283 [hep-ph]].

\bibitem{Herrero:1998eq}
M.~Herrero,
NATO Sci. Ser. C \textbf{534}, 1-59 (1999)
doi:10.1007/978-94-011-4689-0\_1
[arXiv:hep-ph/9812242 [hep-ph]].


\bibitem{Langacker:2009my}
P.~Langacker,
doi:10.1142/9789812838360\_0001
[arXiv:0901.0241 [hep-ph]].


\bibitem{Kibble:2014spa}
T.~W.~B.~Kibble,
[arXiv:1412.4094 [physics.hist-ph]].


\bibitem{Dawson:1994ri}
S.~Dawson,
[arXiv:hep-ph/9411325 [hep-ph]].


\bibitem{Organtini:2012ut}
G.~Organtini,
Eur. J. Phys. \textbf{33}, 1397-1406 (2012)
doi:10.1088/0143-0807/33/5/1397
[arXiv:1207.2146 [physics.pop-ph]].


\bibitem{Peskin:2015kka}
M.~E.~Peskin,
Annalen Phys. \textbf{528}, no.1-2, 20-34 (2016)
doi:10.1002/andp.201500225
[arXiv:1506.08185 [hep-ph]].


\bibitem{Bass:2021acr}
S.~D.~Bass, A.~De Roeck and M.~Kado,
Nature Rev. Phys. \textbf{3}, no.9, 608-624 (2021)
doi:10.1038/s42254-021-00341-2
[arXiv:2104.06821 [hep-ph]].


\bibitem{ATLAS:2012yve}
G.~Aad \textit{et al.} [ATLAS],
Phys. Lett. B \textbf{716}, 1-29 (2012)
doi:10.1016/j.physletb.2012.08.020
[arXiv:1207.7214 [hep-ex]];


\bibitem{CMS:2012qbp}
S.~Chatrchyan \textit{et al.} [CMS],
Phys. Lett. B \textbf{716}, 30-61 (2012)
doi:10.1016/j.physletb.2012.08.021
[arXiv:1207.7235 [hep-ex]].


\bibitem{Apollinari:2015wtw}
G.~Apollinari, O.~Br\"uning, T.~Nakamoto and L.~Rossi,
CERN Yellow Rep., no.5, 1-19 (2015)
doi:10.5170/CERN-2015-005.1
[arXiv:1705.08830 [physics.acc-ph]].


\bibitem{Manousakis:2022ecy}
E.~Manousakis,
Phys. Lett. B \textbf{829}, 137049 (2022)
doi:10.1016/j.physletb.2022.137049
[arXiv:2204.03617 [hep-th]].


\bibitem{Dine:2003ax}
M.~Dine and A.~Kusenko,
Rev. Mod. Phys. \textbf{76}, 1 (2003)
doi:10.1103/RevModPhys.76.1
[arXiv:hep-ph/0303065 [hep-ph]].


\bibitem{Morrissey:2012db}
D.~E.~Morrissey and M.~J.~Ramsey-Musolf,
New J. Phys. \textbf{14}, 125003 (2012)
doi:10.1088/1367-2630/14/12/125003
[arXiv:1206.2942 [hep-ph]].


\bibitem{Buchmuller:2005eh}
W.~Buchmuller, R.~D.~Peccei and T.~Yanagida,
Ann. Rev. Nucl. Part. Sci. \textbf{55}, 311-355 (2005)
doi:10.1146/annurev.nucl.55.090704.151558
[arXiv:hep-ph/0502169 [hep-ph]].


\bibitem{Gildener:1976ih}
E.~Gildener and S.~Weinberg,
Phys. Rev. D \textbf{13}, 3333 (1976)
doi:10.1103/PhysRevD.13.3333.


\bibitem{Weinberg:1975gm}
S.~Weinberg,
Phys. Rev. D \textbf{13}, 974-996 (1976)
doi:10.1103/PhysRevD.19.1277.


\bibitem{Susskind:1978ms}
L.~Susskind,
Phys. Rev. D \textbf{20}, 2619-2625 (1979)
doi:10.1103/PhysRevD.20.2619


\bibitem{Jungman:1995df}
G.~Jungman, M.~Kamionkowski and K.~Griest,
Phys. Rept. \textbf{267}, 195-373 (1996)
doi:10.1016/0370-1573(95)00058-5
[arXiv:hep-ph/9506380 [hep-ph]].


\bibitem{Sadeghian:2013bga}
L.~Sadeghian,
doi:10.7936/K7JM27QG
[arXiv:1308.5378 [gr-qc]].


\bibitem{Gil-Botella:2013bnb}
I.~Gil-Botella,
doi:10.5170/CERN-2013-003.157
[arXiv:1504.03551 [hep-ph]];


\bibitem{Lyth:1998xn}
D.~H.~Lyth and A.~Riotto,
Phys. Rept. \textbf{314}, 1-146 (1999)
doi:10.1016/S0370-1573(98)00128-8
[arXiv:hep-ph/9807278 [hep-ph]].


\bibitem{Sher:1988mj}
M.~Sher,
Phys. Rept. \textbf{179}, 273-418 (1989)
doi:10.1016/0370-1573(89)90061-6.


\bibitem{Degrassi:2012ry}
G.~Degrassi, S.~Di Vita, J.~Elias-Miro, J.~R.~Espinosa, G.~F.~Giudice, G.~Isidori and A.~Strumia,
JHEP \textbf{08}, 098 (2012)
doi:10.1007/JHEP08(2012)098
[arXiv:1205.6497 [hep-ph]].


\bibitem{Zhang:1994fb}
X.~Zhang, S.~K.~Lee, K.~Whisnant and B.~L.~Young,
Phys. Rev. D \textbf{50}, 7042-7047 (1994)
doi:10.1103/PhysRevD.50.7042
[arXiv:hep-ph/9407259 [hep-ph]].


\bibitem{Kobakhidze:2015xlz}
A.~Kobakhidze, L.~Wu and J.~Yue,
JHEP \textbf{04}, 011 (2016)
doi:10.1007/JHEP04(2016)011
[arXiv:1512.08922 [hep-ph]].


\bibitem{Huang:2015izx}
F.~P.~Huang, P.~H.~Gu, P.~F.~Yin, Z.~H.~Yu and X.~Zhang,
Phys. Rev. D \textbf{93}, no.10, 103515 (2016)
doi:10.1103/PhysRevD.93.103515
[arXiv:1511.03969 [hep-ph]].


\bibitem{Ellis:2013lra}
J.~Ellis and T.~You,
JHEP \textbf{06}, 103 (2013)
doi:10.1007/JHEP06(2013)103
[arXiv:1303.3879 [hep-ph]].


\bibitem{Brod:2013cka}
J.~Brod, U.~Haisch and J.~Zupan,
JHEP \textbf{11}, 180 (2013)
doi:10.1007/JHEP11(2013)180
[arXiv:1310.1385 [hep-ph]].


\bibitem{ACME:2013pal}
J.~Baron \textit{et al.} [ACME],
Science \textbf{343}, 269-272 (2014)
doi:10.1126/science.1248213
[arXiv:1310.7534 [physics.atom-ph]].


\bibitem{Goldstein:2000bp}
J.~Goldstein, C.~S.~Hill, J.~Incandela, S.~J.~Parke, D.~L.~Rainwater and D.~Stuart,
Phys. Rev. Lett. \textbf{86}, 1694-1697 (2001)
doi:10.1103/PhysRevLett.86.1694
[arXiv:hep-ph/0006311 [hep-ph]].


\bibitem{Belyaev:2002ua}
A.~Belyaev and L.~Reina,
JHEP \textbf{08}, 041 (2002)
doi:10.1088/1126-6708/2002/08/041
[arXiv:hep-ph/0205270 [hep-ph]].


\bibitem{Drollinger:2001ym}
V.~Drollinger, T.~Muller and D.~Denegri,
[arXiv:hep-ph/0111312 [hep-ph]].


\bibitem{Beenakker:2002nc}
W.~Beenakker, S.~Dittmaier, M.~Kramer, B.~Plumper, M.~Spira and P.~M.~Zerwas,
Nucl. Phys. B \textbf{653}, 151-203 (2003)
doi:10.1016/S0550-3213(03)00044-0
[arXiv:hep-ph/0211352 [hep-ph]].


\bibitem{Agrawal:2013owa}
P.~Agrawal, S.~Bandyopadhyay and S.~P.~Das,
Phys. Rev. D \textbf{88}, no.9, 093008 (2013)
doi:10.1103/PhysRevD.88.093008
[arXiv:1308.3043 [hep-ph]].


\bibitem{Biswas:2014hwa}
S.~Biswas, R.~Frederix, E.~Gabrielli and B.~Mele,
JHEP \textbf{07}, 020 (2014)
doi:10.1007/JHEP07(2014)020
[arXiv:1403.1790 [hep-ph]].


\bibitem{Garzelli:2011vp}
M.~V.~Garzelli, A.~Kardos, C.~G.~Papadopoulos and Z.~Trocsanyi,
EPL \textbf{96}, no.1, 11001 (2011)
doi:10.1209/0295-5075/96/11001
[arXiv:1108.0387 [hep-ph]].


\bibitem{Frederix:2011zi}
R.~Frederix, S.~Frixione, V.~Hirschi, F.~Maltoni, R.~Pittau and P.~Torrielli,
Phys. Lett. B \textbf{701}, 427-433 (2011)
doi:10.1016/j.physletb.2011.06.012
[arXiv:1104.5613 [hep-ph]].


\bibitem{Degrande:2012gr}
C.~Degrande, J.~M.~Gerard, C.~Grojean, F.~Maltoni and G.~Servant,
JHEP \textbf{07}, 036 (2012)
[erratum: JHEP \textbf{03}, 032 (2013)]
doi:10.1007/JHEP07(2012)036
[arXiv:1205.1065 [hep-ph]].


\bibitem{Curtin:2013zua}
D.~Curtin, J.~Galloway and J.~G.~Wacker,
Phys. Rev. D \textbf{88}, no.9, 093006 (2013)
doi:10.1103/PhysRevD.88.093006
[arXiv:1306.5695 [hep-ph]].


\bibitem{Adelman:2013vro}
J.~Adelman, A.~Loginov, P.~Tipton and J.~Vasquez,
[arXiv:1310.1132 [hep-ex]].


\bibitem{Marciano:1991qq}
W.~J.~Marciano and F.~E.~Paige,
Phys. Rev. Lett. \textbf{66}, 2433-2435 (1991)
doi:10.1103/PhysRevLett.66.2433.


\bibitem{Buckley:2015vsa}
M.~R.~Buckley and D.~Goncalves,
Phys. Rev. Lett. \textbf{116}, no.9, 091801 (2016)
doi:10.1103/PhysRevLett.116.091801
[arXiv:1507.07926 [hep-ph]].


\bibitem{Cao:2016wib}
Q.~H.~Cao, S.~L.~Chen and Y.~Liu,
Phys. Rev. D \textbf{95}, no.5, 053004 (2017)
doi:10.1103/PhysRevD.95.053004
[arXiv:1602.01934 [hep-ph]].


\bibitem{Maltoni:2016yxb}
F.~Maltoni, E.~Vryonidou and C.~Zhang,
JHEP \textbf{10}, 123 (2016)
doi:10.1007/JHEP10(2016)123
[arXiv:1607.05330 [hep-ph]].


\bibitem{Gritsan:2016hjl}
A.~V.~Gritsan, R.~R\"ontsch, M.~Schulze and M.~Xiao,
Phys. Rev. D \textbf{94}, no.5, 055023 (2016)
doi:10.1103/PhysRevD.94.055023
[arXiv:1606.03107 [hep-ph]].


\bibitem{Chang:2016mso}
J.~Chang, K.~Cheung, J.~S.~Lee and C.~T.~Lu,
JHEP \textbf{04}, 138 (2017)
doi:10.1007/JHEP04(2017)138
[arXiv:1607.06566 [hep-ph]].


\bibitem{Bordes:1992jy}
G.~Bordes and B.~van Eijk,
Phys. Lett. B \textbf{299}, 315-320 (1993)
doi:10.1016/0370-2693(93)90266-K.


\bibitem{Ballestrero:1992bk}
A.~Ballestrero and E.~Maina,
Phys. Lett. B \textbf{299}, 312-314 (1993)
doi:10.1016/0370-2693(93)90265-J.


\bibitem{Stirling:1992fx}
W.~J.~Stirling and D.~J.~Summers,
Phys. Lett. B \textbf{283}, 411-415 (1992)
doi:10.1016/0370-2693(92)90040-B.


\bibitem{Diaz-Cruz:1991bks}
J.~L.~Diaz-Cruz and O.~A.~Sampayo,
Phys. Lett. B \textbf{276}, 211-213 (1992)
doi:10.1016/0370-2693(92)90565-L.


\bibitem{Maltoni:2001hu}
F.~Maltoni, K.~Paul, T.~Stelzer and S.~Willenbrock,
Phys. Rev. D \textbf{64}, 094023 (2001)
doi:10.1103/PhysRevD.64.094023
[arXiv:hep-ph/0106293 [hep-ph]].


\bibitem{Lu:2010zzb}
G.~R.~Lu and L.~Wu,
Chin. Phys. Lett. \textbf{27}, 031401 (2010)
doi:10.1088/0256-307X/27/3/031401.


\bibitem{Farina:2012xp}
M.~Farina, C.~Grojean, F.~Maltoni, E.~Salvioni and A.~Thamm,
JHEP \textbf{05}, 022 (2013)
doi:10.1007/JHEP05(2013)022
[arXiv:1211.3736 [hep-ph]].


\bibitem{Biswas:2012bd}
S.~Biswas, E.~Gabrielli and B.~Mele,
JHEP \textbf{01}, 088 (2013)
doi:10.1007/JHEP01(2013)088
[arXiv:1211.0499 [hep-ph]].

\bibitem{Ellis:2013yxa}
J.~Ellis, D.~S.~Hwang, K.~Sakurai and M.~Takeuchi,
JHEP \textbf{04}, 004 (2014)
doi:10.1007/JHEP04(2014)004
[arXiv:1312.5736 [hep-ph]].

\bibitem{Englert:2014pja}
C.~Englert and E.~Re,
Phys. Rev. D \textbf{89}, no.7, 073020 (2014)
doi:10.1103/PhysRevD.89.073020
[arXiv:1402.0445 [hep-ph]].


\bibitem{Chang:2014rfa}
J.~Chang, K.~Cheung, J.~S.~Lee and C.~T.~Lu,
JHEP \textbf{05}, 062 (2014)
doi:10.1007/JHEP05(2014)062
[arXiv:1403.2053 [hep-ph]].

\bibitem{Wu:2014dba}
L.~Wu,
JHEP \textbf{02}, 061 (2015)
doi:10.1007/JHEP02(2015)061
[arXiv:1407.6113 [hep-ph]].


\bibitem{Yang:2014xma}
B.~Yang, J.~Han and N.~Liu,
JHEP \textbf{04}, 148 (2015)
doi:10.1007/JHEP04(2015)148
[arXiv:1412.2927 [hep-ph]].


\bibitem{Yue:2014tya}
J.~Yue,
Phys. Lett. B \textbf{744}, 131-136 (2015)
doi:10.1016/j.physletb.2015.03.044
[arXiv:1410.2701 [hep-ph]].


\bibitem{Rindani:2016scj}
S.~D.~Rindani, P.~Sharma and A.~Shivaji,
Phys. Lett. B \textbf{761}, 25-30 (2016)
doi:10.1016/j.physletb.2016.08.002
[arXiv:1605.03806 [hep-ph]].


\bibitem{Liu:2016gsi}
Y.~B.~Liu and Z.~J.~Xiao,
Phys. Lett. B \textbf{763}, 458-464 (2016)
doi:10.1016/j.physletb.2016.11.004
[arXiv:1610.03250 [hep-ph]].


\bibitem{Kobakhidze:2014gqa}
A.~Kobakhidze, L.~Wu and J.~Yue,
JHEP \textbf{10}, 100 (2014)
doi:10.1007/JHEP10(2014)100
[arXiv:1406.1961 [hep-ph]].


\bibitem{Bortolato:2020zcg}
B.~Bortolato, J.~F.~Kamenik, N.~Ko\v{s}nik and A.~Smolkovi\v{c},
Nucl. Phys. B \textbf{964}, 115328 (2021)
doi:10.1016/j.nuclphysb.2021.115328
[arXiv:2006.13110 [hep-ph]].


\bibitem{Cao:2020hhb}
Q.~H.~Cao, K.~P.~Xie, H.~Zhang and R.~Zhang,
Chin. Phys. C \textbf{45}, no.2, 023117 (2021)
doi:10.1088/1674-1137/abcfac
[arXiv:2008.13442 [hep-ph]].


\bibitem{Goncalves:2021dcu}
D.~Gon\c{c}alves, J.~H.~Kim, K.~Kong and Y.~Wu,
JHEP \textbf{01}, 158 (2022)
doi:10.1007/JHEP01(2022)158
[arXiv:2108.01083 [hep-ph]].


\bibitem{Mileo:2016mxg}
N.~Mileo, K.~Kiers, A.~Szynkman, D.~Crane and E.~Gegner,
JHEP \textbf{07}, 056 (2016)
doi:10.1007/JHEP07(2016)056
[arXiv:1603.03632 [hep-ph]].


\bibitem{Dawson:2003zu}
S.~Dawson, C.~Jackson, L.~H.~Orr, L.~Reina and D.~Wackeroth,
Phys. Rev. D \textbf{68}, 034022 (2003)
doi:10.1103/PhysRevD.68.034022
[arXiv:hep-ph/0305087 [hep-ph]].


\bibitem{Maltoni:2002jr}
F.~Maltoni, D.~L.~Rainwater and S.~Willenbrock,
Phys. Rev. D \textbf{66}, 034022 (2002)
doi:10.1103/PhysRevD.66.034022
[arXiv:hep-ph/0202205 [hep-ph]].


\bibitem{CMS:2018uxb}
A.~M.~Sirunyan \textit{et al.} [CMS],
Phys. Rev. Lett. \textbf{120}, no.23, 231801 (2018)
doi:10.1103/PhysRevLett.120.231801
[arXiv:1804.02610 [hep-ex]].


\bibitem{ATLAS:2018mme}
M.~Aaboud \textit{et al.} [ATLAS],
Phys. Lett. B \textbf{784}, 173-191 (2018)
doi:10.1016/j.physletb.2018.07.035
[arXiv:1806.00425 [hep-ex]].


\bibitem{Khatibi:2014bsa}
S.~Khatibi and M.~Mohammadi Najafabadi,
Phys. Rev. D \textbf{90}, no.7, 074014 (2014)
doi:10.1103/PhysRevD.90.074014
[arXiv:1409.6553 [hep-ph]].


\bibitem{Bahl:2021dnc}
H.~Bahl and S.~Brass,
JHEP \textbf{03}, 017 (2022)
doi:10.1007/JHEP03(2022)017
[arXiv:2110.10177 [hep-ph]].


\bibitem{Schlaffer:2014osa}
M.~Schlaffer, M.~Spannowsky, M.~Takeuchi, A.~Weiler and C.~Wymant,
Eur. Phys. J. C \textbf{74}, no.10, 3120 (2014)
doi:10.1140/epjc/s10052-014-3120-z
[arXiv:1405.4295 [hep-ph]].


\bibitem{CMS:2020dkv}
 [CMS],
CMS-PAS-HIG-19-009.


\bibitem{CMS:2020cga}
A.~M.~Sirunyan \textit{et al.} [CMS],
Phys. Rev. Lett. \textbf{125}, no.6, 061801 (2020)
doi:10.1103/PhysRevLett.125.061801
[arXiv:2003.10866 [hep-ex]].


\bibitem{CMS:2021nnc}
A.~M.~Sirunyan \textit{et al.} [CMS],
Phys. Rev. D \textbf{104}, no.5, 052004 (2021)
doi:10.1103/PhysRevD.104.052004
[arXiv:2104.12152 [hep-ex]].


\bibitem{Nishiwaki:2013cma}
K.~Nishiwaki, S.~Niyogi and A.~Shivaji,
JHEP \textbf{04}, 011 (2014)
doi:10.1007/JHEP04(2014)011
[arXiv:1309.6907 [hep-ph]].


\bibitem{Artoisenet:2013vfa}
P.~Artoisenet, P.~de Aquino, F.~Maltoni and O.~Mattelaer,
Phys. Rev. Lett. \textbf{111}, no.9, 091802 (2013)
doi:10.1103/PhysRevLett.111.091802
[arXiv:1304.6414 [hep-ph]].


\bibitem{Dai:1993gm}
J.~Dai, J.~F.~Gunion and R.~Vega,
Phys. Rev. Lett. \textbf{71}, 2699-2702 (1993)
doi:10.1103/PhysRevLett.71.2699
[arXiv:hep-ph/9306271 [hep-ph]].


\bibitem{Schmidt:1992et}
C.~R.~Schmidt and M.~E.~Peskin,
Phys. Rev. Lett. \textbf{69}, 410-413 (1992)
doi:10.1103/PhysRevLett.69.410


\bibitem{Barman:2021yfh}
R.~K.~Barman, D.~Gon\c{c}alves and F.~Kling,
Phys. Rev. D \textbf{105}, no.3, 035023 (2022)
doi:10.1103/PhysRevD.105.035023
[arXiv:2110.07635 [hep-ph]].


\bibitem{ATLAS:2020ior}
G.~Aad \textit{et al.} [ATLAS],
Phys. Rev. Lett. \textbf{125}, no.6, 061802 (2020)
doi:10.1103/PhysRevLett.125.061802
[arXiv:2004.04545 [hep-ex]].


\bibitem{Bahl:2020wee}
H.~Bahl, P.~Bechtle, S.~Heinemeyer, J.~Katzy, T.~Klingl, K.~Peters, M.~Saimpert, T.~Stefaniak and G.~Weiglein,
JHEP \textbf{11}, 127 (2020)
doi:10.1007/JHEP11(2020)127
[arXiv:2007.08542 [hep-ph]].


\bibitem{Bahl:2022yrs}
H.~Bahl, E.~Fuchs, S.~Heinemeyer, J.~Katzy, M.~Menen, K.~Peters, M.~Saimpert and G.~Weiglein,
Eur. Phys. J. C \textbf{82}, no.7, 604 (2022)
doi:10.1140/epjc/s10052-022-10528-1
[arXiv:2202.11753 [hep-ph]].


\bibitem{ATLAS:2023cbt}
 [ATLAS],
[arXiv:2303.05974 [hep-ex]].


\bibitem{Hermann:2022vit}
J.~Hermann, D.~Stremmer and M.~Worek,
JHEP \textbf{09}, 138 (2022)
doi:10.1007/JHEP09(2022)138
[arXiv:2205.09983 [hep-ph]].


\bibitem{Maltoni:2013sma}
F.~Maltoni, K.~Mawatari and M.~Zaro,
Eur. Phys. J. C \textbf{74}, no.1, 2710 (2014)
doi:10.1140/epjc/s10052-013-2710-5
[arXiv:1311.1829 [hep-ph]].


\bibitem{Demartin:2014fia}
F.~Demartin, F.~Maltoni, K.~Mawatari, B.~Page and M.~Zaro,
Eur. Phys. J. C \textbf{74}, no.9, 3065 (2014)
doi:10.1140/epjc/s10052-014-3065-2
[arXiv:1407.5089 [hep-ph]].



\bibitem{ATLAS:2021pkb}
G.~Aad \textit{et al.} [ATLAS],
Eur. Phys. J. C \textbf{82}, no.7, 622 (2022)
doi:10.1140/epjc/s10052-022-10366-1
[arXiv:2109.13808 [hep-ex]].


\bibitem{ATLAS:2015xst}
G.~Aad \textit{et al.} [ATLAS],
JHEP \textbf{04}, 117 (2015)
doi:10.1007/JHEP04(2015)117
[arXiv:1501.04943 [hep-ex]].

\bibitem{CMS:2017zyp}
A.~M.~Sirunyan \textit{et al.} [CMS],
Phys. Lett. B \textbf{779}, 283-316 (2018)
doi:10.1016/j.physletb.2018.02.004
[arXiv:1708.00373 [hep-ex]].


\bibitem{CMS:2020rpr}
 [CMS],
CMS-PAS-HIG-20-006.


\bibitem{CMS:2021sdq}
A.~Tumasyan \textit{et al.} [CMS],
JHEP \textbf{06}, 012 (2022)
doi:10.1007/JHEP06(2022)012
[arXiv:2110.04836 [hep-ex]].


\bibitem{ATLAS:2020fzp}
G.~Aad \textit{et al.} [ATLAS],
Phys. Lett. B \textbf{812}, 135980 (2021)
doi:10.1016/j.physletb.2020.135980
[arXiv:2007.07830 [hep-ex]].


\bibitem{CMS:2020xwi}
A.~M.~Sirunyan \textit{et al.} [CMS],
JHEP \textbf{01}, 148 (2021)
doi:10.1007/JHEP01(2021)148
[arXiv:2009.04363 [hep-ex]].


\bibitem{Cirigliano:2016nyn}
V.~Cirigliano, W.~Dekens, J.~de Vries and E.~Mereghetti,
Phys. Rev. D \textbf{94}, no.3, 034031 (2016)
doi:10.1103/PhysRevD.94.034031
[arXiv:1605.04311 [hep-ph]].


\bibitem{Gupta:2009wu}
S.~K.~Gupta, A.~S.~Mete and G.~Valencia,
Phys. Rev. D \textbf{80}, 034013 (2009)
doi:10.1103/PhysRevD.80.034013
[arXiv:0905.1074 [hep-ph]].


\bibitem{Tiwari:2019kly}
A.~Tiwari and S.~Kumar Gupta,
Adv. High Energy Phys. \textbf{2021}, 6676930 (2021)
doi:10.1155/2021/6676930
[arXiv:1903.05365 [hep-ph]].


\bibitem{Alioli:2017ces}
S.~Alioli, V.~Cirigliano, W.~Dekens, J.~de Vries and E.~Mereghetti,
JHEP \textbf{05}, 086 (2017)
doi:10.1007/JHEP05(2017)086
[arXiv:1703.04751 [hep-ph]].


\bibitem{Aguilar-Saavedra:2008quj}
J.~A.~Aguilar-Saavedra,
Nucl. Phys. B \textbf{804}, 160-192 (2008)
doi:10.1016/j.nuclphysb.2008.06.013
[arXiv:0803.3810 [hep-ph]].


\bibitem{Tiwari:2022nli}
A.~Tiwari and S.~K.~Gupta,
Nucl. Phys. B \textbf{982}, 115898 (2022)
doi:10.1016/j.nuclphysb.2022.115898
[arXiv:2204.12800 [hep-ph]].


\bibitem{CMS:2014uod}
V.~Khachatryan \textit{et al.} [CMS],
JHEP \textbf{01}, 053 (2015)
doi:10.1007/JHEP01(2015)053
[arXiv:1410.1154 [hep-ex]].


\bibitem{Gupta:2009eq}
S.~K.~Gupta and G.~Valencia,
Phys. Rev. D \textbf{81}, 034013 (2010)
doi:10.1103/PhysRevD.81.034013
[arXiv:0912.0707 [hep-ph]].


\bibitem{Dolan:2014upa}
M.~J.~Dolan, P.~Harris, M.~Jankowiak and M.~Spannowsky,
Phys. Rev. D \textbf{90}, 073008 (2014)
doi:10.1103/PhysRevD.90.073008
[arXiv:1406.3322 [hep-ph]].


\bibitem{Englert:2012xt}
C.~Englert, D.~Goncalves-Netto, K.~Mawatari and T.~Plehn,
JHEP \textbf{01}, 148 (2013)
doi:10.1007/JHEP01(2013)148
[arXiv:1212.0843 [hep-ph]].


\bibitem{Kobakhidze:2016mfx}
A.~Kobakhidze, N.~Liu, L.~Wu and J.~Yue,
Phys. Rev. D \textbf{95}, no.1, 015016 (2017)
doi:10.1103/PhysRevD.95.015016
[arXiv:1610.06676 [hep-ph]].


\bibitem{Bernlochner:2018opw}
F.~U.~Bernlochner, C.~Englert, C.~Hays, K.~Lohwasser, H.~Mildner, A.~Pilkington, D.~D.~Price and M.~Spannowsky,
Phys. Lett. B \textbf{790}, 372-379 (2019)
doi:10.1016/j.physletb.2019.01.043
[arXiv:1808.06577 [hep-ph]].


\bibitem{Englert:2019xhk}
C.~Englert, P.~Galler, A.~Pilkington and M.~Spannowsky,
Phys. Rev. D \textbf{99}, no.9, 095007 (2019)
doi:10.1103/PhysRevD.99.095007
[arXiv:1901.05982 [hep-ph]].


\bibitem{Gritsan:2020pib}
A.~V.~Gritsan, J.~Roskes, U.~Sarica, M.~Schulze, M.~Xiao and Y.~Zhou,
Phys. Rev. D \textbf{102}, no.5, 056022 (2020)
doi:10.1103/PhysRevD.102.056022
[arXiv:2002.09888 [hep-ph]].



























\bibitem{Accomando:2006ga}
E.~Accomando, A.~G.~Akeroyd, E.~Akhmetzyanova, J.~Albert, A.~Alves, N.~Amapane, M.~Aoki, G.~Azuelos, S.~Baffioni and A.~Ballestrero, \textit{et al.}
doi:10.5170/CERN-2006-009
[arXiv:hep-ph/0608079 [hep-ph]].


\bibitem{Koulovassilopoulos:1993pw}
V.~Koulovassilopoulos and R.~S.~Chivukula,
Phys. Rev. D \textbf{50}, 3218-3234 (1994)
doi:10.1103/PhysRevD.50.3218
[arXiv:hep-ph/9312317 [hep-ph]].


\bibitem{CMS:2014nkk}
V.~Khachatryan \textit{et al.} [CMS],
Phys. Rev. D \textbf{92}, no.1, 012004 (2015)
doi:10.1103/PhysRevD.92.012004
[arXiv:1411.3441 [hep-ex]].


\bibitem{CMS:2016tad}
V.~Khachatryan \textit{et al.} [CMS],
Phys. Lett. B \textbf{759}, 672-696 (2016)
doi:10.1016/j.physletb.2016.06.004
[arXiv:1602.04305 [hep-ex]].


\bibitem{Binosi:2003yf}
D.~Binosi and L.~Theussl,
Comput. Phys. Commun. \textbf{161}, 76-86 (2004)
doi:10.1016/j.cpc.2004.05.001
[arXiv:hep-ph/0309015 [hep-ph]].


\bibitem{Binosi:2008ig}
D.~Binosi, J.~Collins, C.~Kaufhold and L.~Theussl,
Comput. Phys. Commun. \textbf{180}, 1709-1715 (2009)
doi:10.1016/j.cpc.2009.02.020
[arXiv:0811.4113 [hep-ph]].


\bibitem{ParticleDataGroup:2022pth}
R.~L.~Workman \textit{et al.} [Particle Data Group],
PTEP \textbf{2022}, 083C01 (2022)
doi:10.1093/ptep/ptac097


\bibitem{Demartin:2015uha}
F.~Demartin, F.~Maltoni, K.~Mawatari and M.~Zaro,
Eur. Phys. J. C \textbf{75}, no.6, 267 (2015)
doi:10.1140/epjc/s10052-015-3475-9
[arXiv:1504.00611 [hep-ph]].


\bibitem{Goncalves:2018agy}
D.~Gon\c{c}alves, K.~Kong and J.~H.~Kim,
JHEP \textbf{06}, 079 (2018)
doi:10.1007/JHEP06(2018)079
[arXiv:1804.05874 [hep-ph]].


\bibitem{Gunion:1996xu}
J.~F.~Gunion and X.~G.~He,
Phys. Rev. Lett. \textbf{76}, 4468-4471 (1996)
doi:10.1103/PhysRevLett.76.4468
[arXiv:hep-ph/9602226 [hep-ph]].


\bibitem{Hayreter:2015ryk}
A.~Hayreter and G.~Valencia,
Phys. Rev. D \textbf{93}, no.1, 014020 (2016)
doi:10.1103/PhysRevD.93.014020
[arXiv:1511.01464 [hep-ph]].


\bibitem{Gupta:2009pv}
S.~K.~Gupta,
[arXiv:0910.0068 [hep-ph]].


\bibitem{Han:2009ra}
T.~Han and Y.~Li,
Phys. Lett. B \textbf{683}, 278-281 (2010)
doi:10.1016/j.physletb.2009.12.047
[arXiv:0911.2933 [hep-ph]].


\bibitem{Artoisenet:2013puc}
P.~Artoisenet, P.~de Aquino, F.~Demartin, R.~Frederix, S.~Frixione, F.~Maltoni, M.~K.~Mandal, P.~Mathews, K.~Mawatari and V.~Ravindran, \textit{et al.}
JHEP \textbf{11}, 043 (2013)
doi:10.1007/JHEP11(2013)043
[arXiv:1306.6464 [hep-ph]].


\bibitem{Alloul:2013bka}
A.~Alloul, N.~D.~Christensen, C.~Degrande, C.~Duhr and B.~Fuks,
Comput. Phys. Commun. \textbf{185}, 2250-2300 (2014)
doi:10.1016/j.cpc.2014.04.012
[arXiv:1310.1921 [hep-ph]].


\bibitem{Christensen:2008py}
N.~D.~Christensen and C.~Duhr,
Comput. Phys. Commun. \textbf{180}, 1614-1641 (2009)
doi:10.1016/j.cpc.2009.02.018
[arXiv:0806.4194 [hep-ph]].


\bibitem{Alwall:2011uj}
J.~Alwall, M.~Herquet, F.~Maltoni, O.~Mattelaer and T.~Stelzer,
JHEP \textbf{06}, 128 (2011)
doi:10.1007/JHEP06(2011)128
[arXiv:1106.0522 [hep-ph]].


\bibitem{Frederix:2009yq}
R.~Frederix, S.~Frixione, F.~Maltoni and T.~Stelzer,
JHEP \textbf{10}, 003 (2009)
doi:10.1088/1126-6708/2009/10/003
[arXiv:0908.4272 [hep-ph]].


\bibitem{Alwall:2014hca}
J.~Alwall, R.~Frederix, S.~Frixione, V.~Hirschi, F.~Maltoni, O.~Mattelaer, H.~S.~Shao, T.~Stelzer, P.~Torrielli and M.~Zaro,
JHEP \textbf{07}, 079 (2014)
doi:10.1007/JHEP07(2014)079
[arXiv:1405.0301 [hep-ph]].


\bibitem{Hirschi:2011pa}
V.~Hirschi, R.~Frederix, S.~Frixione, M.~V.~Garzelli, F.~Maltoni and R.~Pittau,
JHEP \textbf{05}, 044 (2011)
doi:10.1007/JHEP05(2011)044
[arXiv:1103.0621 [hep-ph]].


\bibitem{Artoisenet:2012st}
P.~Artoisenet, R.~Frederix, O.~Mattelaer and R.~Rietkerk,
JHEP \textbf{03}, 015 (2013)
doi:10.1007/JHEP03(2013)015
[arXiv:1212.3460 [hep-ph]].


\bibitem{Sjostrand:2014zea}
T.~Sj\"ostrand, S.~Ask, J.~R.~Christiansen, R.~Corke, N.~Desai, P.~Ilten, S.~Mrenna, S.~Prestel, C.~O.~Rasmussen and P.~Z.~Skands,
Comput. Phys. Commun. \textbf{191}, 159-177 (2015)
doi:10.1016/j.cpc.2015.01.024
[arXiv:1410.3012 [hep-ph]].


\bibitem{Bierlich:2022pfr}
C.~Bierlich, S.~Chakraborty, N.~Desai, L.~Gellersen, I.~Helenius, P.~Ilten, L.~L\"onnblad, S.~Mrenna, S.~Prestel and C.~T.~Preuss, \textit{et al.}
doi:10.21468/SciPostPhysCodeb.8
[arXiv:2203.11601 [hep-ph]].


\bibitem{Ball:2013hta}
R.~D.~Ball \textit{et al.} [NNPDF],
Nucl. Phys. B \textbf{877}, 290-320 (2013)
doi:10.1016/j.nuclphysb.2013.10.010
[arXiv:1308.0598 [hep-ph]].


\bibitem{NNPDF:2014otw}
R.~D.~Ball \textit{et al.} [NNPDF],
JHEP \textbf{04}, 040 (2015)
doi:10.1007/JHEP04(2015)040
[arXiv:1410.8849 [hep-ph]].


\bibitem{CMS:2013szn}
S.~Chatrchyan \textit{et al.} [CMS],
JHEP \textbf{05}, 145 (2013)
doi:10.1007/JHEP05(2013)145
[arXiv:1303.0763 [hep-ex]].


\bibitem{Azevedo:2022jnd}
D.~Azevedo, R.~Capucha, A.~Onofre and R.~Santos,
JHEP \textbf{09}, 246 (2022)
doi:10.1007/JHEP09(2022)246
[arXiv:2208.04271 [hep-ph]].


\bibitem{Barman:2022pip}
R.~K.~Barman, M.~E.~Cassidy, Z.~Dong, D.~Gon\c{c}alves, J.~H.~Kim, F.~Kling, K.~Kong, I.~M.~Lewis, Y.~Wu and Y.~Zhang, \textit{et al.}
[arXiv:2203.08127 [hep-ph]].

\end{thebibliography}
\end{document}